\documentclass[preprint,12pt,3p]{elsarticle}

\usepackage{amssymb}
\usepackage{amsmath}
\usepackage{amsthm}
\usepackage{graphicx}    
\usepackage{subcaption}  
\usepackage{booktabs}
\usepackage{enumitem}
\usepackage{lmodern}
\usepackage{url}

\journal{pre-print}

\begin{document}

\begin{frontmatter}



\title{Learned Warm Starts for Steady Transonic RANS Simulations \\ of Wing and Aircraft Configurations} 


\author[authors]{Muhammad Irfan Zafar}\ead{muhammadirfan.zafar@tai.com.tr}\cortext[]{Corresponding author} \author[authors]{Saleh Abuhanieh} \author[authors]{\c{S}ahin Yi\u{g}it} \affiliation[authors]{organization={Computational Fluid Mechanics Department}, addressline={Turkish Aerospace Industries}, city={Ankara}, country={Turkey}}

\begin{abstract}
Neural surrogates can accelerate many--query aerodynamic analyses, but their standalone predictions may be insufficiently accurate or robust to support reliable design evaluations. We develop a learned warm--start formulation that predicts mean--flow and turbulence fields directly on native CFD meshes for subsequent correction by the RANS solver.  A common architecture and objective structure are used across subcritical and transonic flows, including challenging conditions with shock--induced separation. A condition--modulated Transolver combines Fourier and wall features with target--gated regularization, a Sobolev--type loss, adaptive loss weighting and sampled training. Prediction requires no preliminary flow solution, and training requires no solver--residual evaluation.  Extensive three--dimensional CFD evaluations on SuperWing, the NASA Common Research Model (CRM) and DLR--F6 assess geometric and operating--condition generalization, including interpolation, extrapolation and cross--mesh deployment, using Spalart--Allmaras and $k$--$\omega$ SST closures on million--cell meshes. Unseen nominal cases achieve geometric--mean iteration speed--ups of approximately $2$--$4\times$ under a common convergence criterion based on cold--run aerodynamic coefficients. On CRM, learned initialization provides greater aggregate acceleration than nearest--condition restarts using the same solution library. Resolution invariance enables direct deployment of coarse--mesh models on finer meshes without retraining, solution interpolation or target--condition coarse solves. Including data generation and training, estimated campaign speed--ups reach $2.1$--$2.3\times$. No warm run diverges; fewer than $2\%$ fail to satisfy the cold--reference convergence criterion. These results support learned initialization for repeated aerodynamic analyses, while highlighting initialization sensitivity near separation as a remaining limitation.
\end{abstract}




\end{frontmatter}



\section{Introduction}
\label{sec1}
Computational fluid dynamics (CFD) enables aerodynamic analysis by numerically solving the partial differential equations (PDEs) governing fluid motion. At flight Reynolds numbers, steady Reynolds--averaged Navier--Stokes (RANS) modeling remains an industrial workhorse~\citep{SlotnickEtAl2014CFDVision2030,SpalartVenkatakrishnan2016}. Large simulations may require thousands of iterations to reduce discrete residuals to tolerance level across millions of unknowns~\citep{YildirimEtAl2019}. In the many--query settings, which dominate the design phase, this compute cost accumulates during shape optimization, aerodynamic loads database generation, and uncertainty quantification. These tasks repeatedly evaluate geometric configurations and operating flow conditions that often differ only slightly.

\subsection*{Aerodynamic Surrogate Modeling}
Neural surrogates exploit this opportunity by learning solution maps from precomputed flow fields~\citep{LuEtAl2021DeepONet,LiEtAl2021FNO}. Their architectures have progressively accommodated more complex geometries and larger discretizations. Convolutional networks predict fields on regular or mapped grids~\citep{GuoEtAl2016,ThuereyEtAl2020}. Graph networks operate on simulation meshes, while neural operators learn mappings between input and solution functions~\citep{PfaffEtAl2021,LuEtAl2021DeepONet,LiEtAl2021FNO,zafar2022frame}. Geometry--aware operators extend these mappings to irregular domains~\citep{LiEtAl2023GeoFNO,LiEtAl2023GINO}. Recent frameworks, including Transolver, AB-UPT, and DoMINO, address large three--dimensional discretizations~\citep{ZhouEtAl2026Transolver3,AlkinEtAl2025ABUPT,RanadeEtAl2025DoMINO}. AeroJEPA predicts compact latent flow representations and reconstructs fields through an implicit decoder, separating latent prediction from output resolution~\citep{GiralEtAl2026AeroJEPA}.


However, computational scale and domain--averaged accuracy alone do not establish surrogate reliability. That distinction is sharpest in transonic flow, where aggregate metrics can conceal localized errors in shocks, boundary layers, and separated wakes. Predictions that may support preliminary design exploration may still lack the reliability required for loads--database generation and certification. These limitations motivate a complementary role for surrogates: supplying initial fields that conventional solvers correct to prescribed convergence criteria~\citep{ZhouEtAl2025SuperFidelity}.

\subsection*{Learned Warm Starts}
Warm start approach uses a predicted field as an initial iterate while preserving the target discrete equations. Its objective is to produce an initial state that the solver can correct efficiently. Predictive accuracy alone does not establish this property. In nonlinear elasticity for solids, \citet{OhEtAl2026} showed that small field errors can coexist with an indefinite discrete Jacobian. Although the governing problem differs, this finding illustrates the distinction between accurate prediction and reliable nonlinear initialization. For RANS, the value of a learned initial field must therefore be established through its effect on numerical convergence.

Existing RANS studies demonstrate both the potential of learned initialization and its dependence on the available inputs. CFDNet uses intermediate solver states within a warm--up and refinement procedure on structured grids~\citep{ObiolsSalesEtAl2020CFDNet}. It therefore requires preliminary iterations rather than predicting solely from geometry and operating conditions. The super--fidelity framework presented by \citet{ZhouEtAl2025SuperFidelity} maps potential--flow solutions to RANS fields on irregular discretizations. This supplies physical information but adds a low--fidelity solve to the initialization pipeline. \citet{FuchiEtAl2025} developed a CNN wake-extension model to accelerate incompressible RANS simulations around a NACA0012 airfoil. Their study highlighted the benefit of representing the wake to the downstream boundary. However, the near--body field was supplied from a converged solution, leaving the effect of near--body prediction errors untested. For transient automotive flows, \citet{SharpeEtAl2025} accelerated the approach to statistical steadiness. That objective differs from convergence of a steady RANS residual.

Aircraft–scale studies reveal further challenges in isolating the benefit of initialization. \citet{LiuEtAl2026MHLF} predicted coarse–grid mean–flow fields using region–specific modeling and prolonged them to fine structured multiblock meshes. Turbulence variables were supplied from coarse–grid solutions at nearby angles of attack. Subsequent CFD correction updated the boundary–layer region more frequently than the inner and outer regions, whereas the cold–start baseline used synchronous updates throughout the domain. Their reported acceleration therefore combines learned initialization with a modified iteration schedule; it does not isolate the contribution of initialization alone. The authors also reported that different initializations could produce different final aerodynamic coefficients, but did not specify how many cases were affected. Several comparison point–cloud models increased computational cost or failed during CFD correction, underscoring the need to assess both convergence efficiency and agreement with the reference solution when evaluating learned warm starts.
\citet{LeiEtAl2026} used surrogate predictions to initialize Newton--Krylov correction of RANS--SA solutions in ADflow. Their two--dimensional training incorporates a differentiable residual evaluator, requiring a compatible implementation of the discrete equations when adapting the method to another solver. They also present a three--dimensional flying--wing extension with residual--reduction and solver--time results, although explicit three-dimensional success/failure counts and per--case speed--up statistics are not reported. The reported validation uses fixed-topology structured meshes; performance on unstructured meshes with varying connectivity remains unassessed.

\subsection*{Challenges and Contributions}
A practical RANS warm--start model must provide physically admissible mean--flow and turbulence fields while preserving shocks, boundary layers, and wakes without introducing spurious oscillations. Model training and prediction should remain computationally affordable on large native unstructured meshes and reliable beyond the training conditions. Broader applicability calls for a formulation that accommodates different turbulence closures and solver interfaces without preliminary flow solutions or solver--dependent residual training. Evaluation must establish solver acceleration under matched numerical settings and application-relevant convergence criteria. Agreement between converged solutions requires explicit verification because steady RANS can admit multiple solutions~\citep{KamenetskiyEtAl2014}. Comparisons should include conventional restarts and solver--native initialization~\citep{WangEtAl2026ConvergenceBooster}. Finally, practical benefit depends on whether repeated--use savings offset data--generation, training, inference, and field--transfer costs~\citep{ZhouEtAl2025SuperFidelity}.

The present work develops a warm--start formulation guided by flow physics and the requirements of RANS initialization (Section~\ref{sec:formulation}). Its contributions are:

\begin{itemize}
\item \textbf{Native--mesh prediction of mean--flow and turbulence fields.} 
The model predicts freestream--relative fields directly at CFD cells from geometry, wall features and operating conditions. Closure--specific outputs and solver interfaces retain a common architecture and objective structure. Initialization requires no preliminary flow solution or borrowed turbulence fields, and training requires neither solver differentiation nor evaluation of the target discrete residual.

\item \textbf{A scalable, flow--conditioned representation.} 
We combine Transolver's Physics--Attention operations and Transolver--3's sampled--training strategy \citep{WuEtAl2024Transolver,ZhouEtAl2026Transolver3} with Fourier coordinate features \citep{TancikEtAl2020} and operating--condition modulation through learned scaling, shifting and residual gating \citep{PeeblesXie2023}. These adaptations tailor the representation to effective RANS warm starts while retaining sampled training and global full--mesh inference.

\item \textbf{Resolution invariance and cross--mesh deployment.} 
Shared cell--wise mappings and global slice aggregation allow the same trained parameters to operate across mesh resolutions. We demonstrate coarse--to--fine deployment without fine--mesh training data, solution interpolation or a target--condition coarse solve, reducing data--generation costs while retaining useful solver acceleration.

\item \textbf{A spatial objective tailored to initialization.}
Building on Sobolev training and its application to CFD initialization \citep{CzarneckiEtAl2017,FuchiEtAl2022MultiFidelity}, a discrete Sobolev--type loss matches relative intercell differences. Complementary target--gated penalties suppress spurious freestream perturbations and intercell oscillations, while relaxing regularization near walls and across physical flow structures.

\item \textbf{Case--adaptive regularization and training balance.}
Per--case target--jump statistics set the scale of the smoothness gate and the floor of the relative jump loss. Roughness--aware face sampling emphasizes localized variations, while a gradient--norm controller adjusts auxiliary loss weights toward prescribed shares \citep{ChenEtAl2018GradNorm,WangEtAl2021Gradient}, balancing field fitting, feature matching and noise suppression.
\end{itemize}

We evaluate the formulation using HiSA with Spalart--Allmaras and $k$--$\omega$ SST closures on SuperWing, the NASA Common Research Model (CRM) and DLR--F6, spanning subcritical to shock--separated flow. The assessment distinguishes initial--field accuracy from solver acceleration and checks agreement with cold--run reference loads under matched numerical settings. Section~2 presents the formulation; Section~3 describes the datasets, assessment protocol and results; Section~4 discusses baselines, campaign costs, robustness and scalability; and Section~5 concludes. The appendices provide implementation details and supplementary results.

\section{Warm--Start Formulation}
\label{sec:formulation}
\begin{figure}[t]
\centering
\includegraphics[trim={0 10mm 0  15mm}, clip, width=0.995\textwidth]{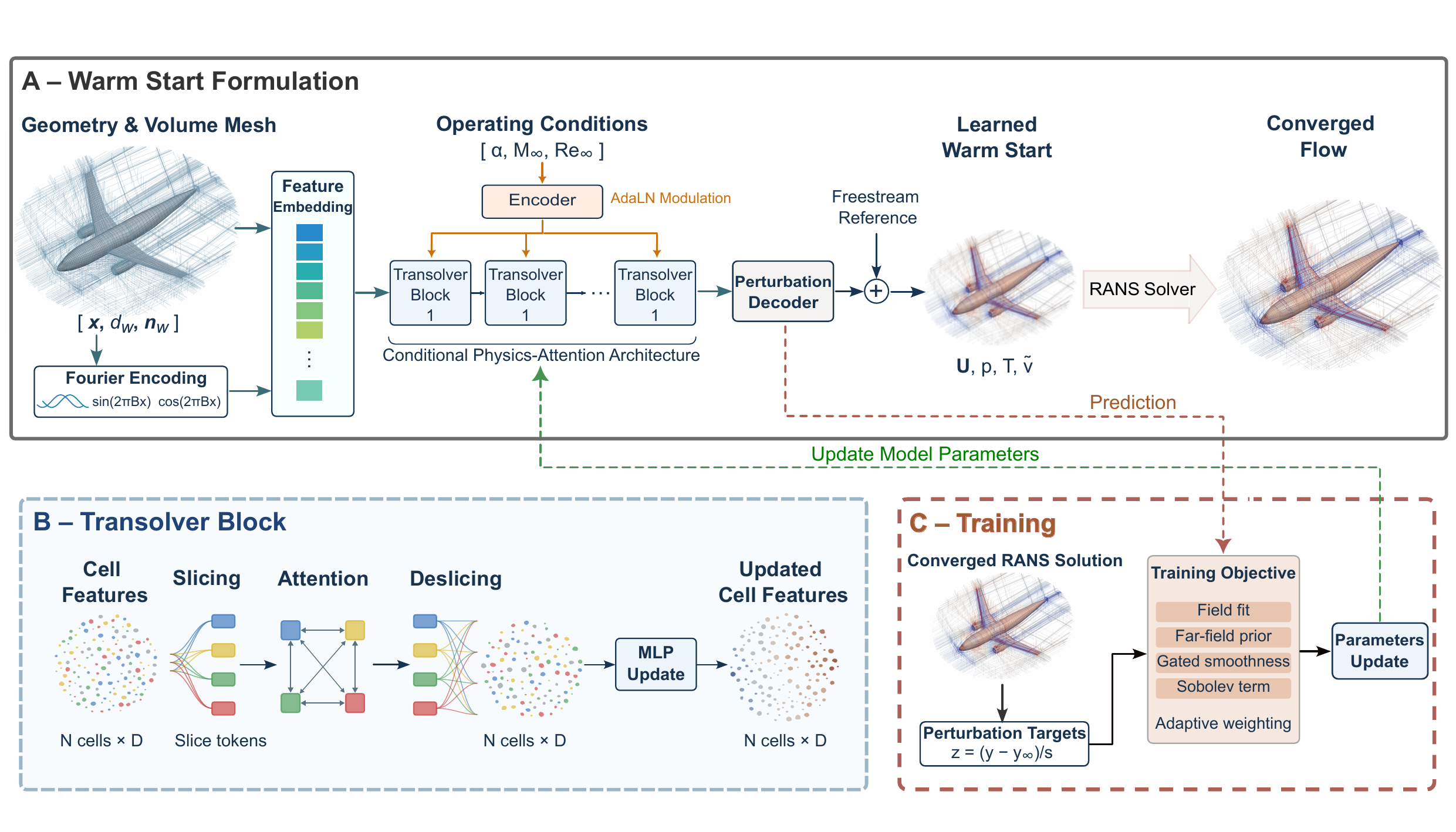}
\caption{Learned warm--start framework for steady Reynolds--averaged Navier--Stokes (RANS) simulations. (A) Fourier--encoded coordinates and wall features are processed by a physics--attention network, conditioned on angle of attack, Mach number and Reynolds number through adaptive layer normalization. Predicted freestream--relative perturbations are reconstructed into mean-flow and turbulence fields that initialize the RANS solver. (B) Each Transolver block aggregates cell features into learned slice tokens, exchanges global information through attention, and redistributes it to cells before a multi--layer perceptron~(MLP) update. (C) Converged RANS fields provide perturbation targets for training, combining field fitting, a far-field prior, gated smoothness and Sobolev--type jump matching with adaptive loss weighting. 
}\label{fig:schematic}
\end{figure}

Figure~\ref{fig:schematic} summarizes the proposed framework, which predicts mean--flow and turbulence fields to initialize a conventional RANS solver. The following subsections describe the flow representation, flow--conditioned architecture, and training objective used to construct these initial fields. Closure--specific output channels and solver--specific restart interfaces connect this common formulation to the selected turbulence model and numerical implementation.

\subsection{Governing problem and learned initialization}
\label{sec:problem}

On a mesh $\mathcal G_h$ with $N$ cells, the discrete steady problem is
\begin{equation}
 \mathcal R_h(\mathbf Q;\boldsymbol\mu,\mathcal B)=\mathbf0,
 \qquad
 \mathbf Q^{(0)}=\mathcal I_{\mathcal B}
 \left[\mathcal D\bigl(\mathcal F_\theta(\mathbf X,\mathbf c)\bigr)\right].
 \label{eq:problem}
\end{equation}
Here $\mathcal R_h$ denotes the discrete residual operator of the steady RANS equations and the selected turbulence closure on mesh $\mathcal G_h$, $\mathbf Q$ is the solver state, $\boldsymbol\mu$ denotes operating conditions, and $\mathcal B$ specifies boundary data.
The neural surrogate model $\mathcal F_\theta$ predicts transformed fields from cell features $\mathbf X$ and conditions $\mathbf c$. The decoder $\mathcal D$ restores dimensional fields, and $\mathcal I_{\mathcal B}$ constructs the restart state. 

The predicted fields initialize the target RANS solver, which then follows its standard iterative solution procedure; model training does not require differentiation through the solver. The predicted mean-flow fields are $\mathbf U$, $p$, and $T$. Turbulence outputs are $\widetilde\nu$ for Spalart--Allmaras (SA), or $(k,\omega)$ for shear-stress transport (SST)~\citep{SpalartAllmaras1992,Menter1994}. Thus the network has six or seven output channels. The restart interface supplies thermodynamic and conservative-variable conversions. The common formulation therefore permits different closures and solvers without requiring identical output dimensions or file formats.

\subsection{Geometry and flow--state representation}
\label{sec:representation}
Each cell receives its coordinates, a compressed dimensionless wall distance, and a unit wall--normal vector. The global operating conditions are angle of attack, Mach number, and logarithmic Reynolds number:
\begin{equation}
 \mathbf x_i=(x_i,y_i,z_i,d_i,\mathbf n_i),\quad
 d_i=\tanh\!\left[\frac{\log_{10}(1+\ell_{w,i}U_\infty/\nu_\infty)}{3}\right],
 \quad \mathbf c=(\alpha,M_\infty,\log_{10}(Re_\infty/10^6)).
 \label{eq:inputs}
\end{equation}
Here $U_\infty=\lVert\mathbf U_\infty\rVert$, $\ell_{w,i}$ is the geometric wall distance of cell $i$, $\nu_\infty$ is the freestream kinematic viscosity, and $Re_\infty=U_\infty L_{\mathrm{ref}}/\nu_\infty$, with reference length $L_{\mathrm{ref}}$. 
Coordinates and operating conditions are standardized componentwise using training--set means and standard deviations ($z$--score standardization). 
The compressed wall--distance and unit--normal components are not standardized. Five Fourier octaves per coordinate add 30 additional input features, with wavelengths based on a fixed training reference length~\citep{TancikEtAl2020}.

The target outputs are dimensionless freestream--relative perturbations of the flow state
$\mathbf y_i=(u_i,v_i,w_i,p_i,T_i,\boldsymbol\ell_{t,i})$:
\begin{equation}
 \mathbf z_i=(\mathbf y_i-\mathbf y_\infty)\oslash\mathbf s,
 \qquad \mathbf s=(U_\infty,U_\infty,U_\infty,q_\infty,T_\infty,\mathbf s_t),
 \quad q_\infty=\tfrac12\rho_\infty U_\infty^2,
 \label{eq:targets}
\end{equation}
where $\oslash$ denotes componentwise division, $\mathbf y_\infty$ contains the corresponding freestream reference values, and $\rho_\infty$ is the freestream density. The pressure target therefore equals the local pressure coefficient, $C_p=(p-p_\infty)/q_\infty$. The vector $\boldsymbol\ell_{t,i}$ contains the transformed turbulence variables, and $\mathbf s_t$ contains their normalization scales. Turbulence variables are represented using logarithmic transforms with positive offsets. Further details for input and output features are specified in \ref{app:features}.

\subsection{Conditional physics-attention architecture}
\label{sec:architecture}
The model maps mesh-derived cell features and operating conditions to normalized flow perturbations using an adapted Transolver backbone~\citep{WuEtAl2024Transolver,LuoEtAl2025TransolverPlusPlus, ZhouEtAl2026Transolver3}. Within each block (Fig.~\ref{fig:schematic}(B)), physics attention aggregates latent cell features into learned slice states, exchanges information between slices, and redistributes the updated information to cells. A cell--wise multilayer perceptron (MLP) then updates the features. This structure enables information exchange across the mesh without pairwise attention over all cells.

An operating-condition encoder conditions both processing stages (physics attention and MLP update) through adaptive layer normalization (AdaLN) and gated residual updates~\citep{PeeblesXie2023}. The operating-condition encoder allows both the physics-attention and MLP stages to adapt to the flow conditions: AdaLN rescales and shifts their normalized input features, while condition-dependent gates control the updates added to the existing features. A perturbation decoder maps the final cell features to the normalized target channels. The network is initialized to predict approximately zero perturbations, corresponding to a near-uniform freestream initialization of the solver's internal fields.

Mesh connectivity is not an input to the network. Native owner--neighbor connectivity is used for face sampling and evaluation of the mesh-based training losses.~\ref{app:architecture} specifies the network configuration, conditioning operations, and optimization settings.

\subsection{Initialization--Oriented Training Objective}
\label{sec:objective}
The training objective shown in Fig.~\ref{fig:schematic}(C) combines field accuracy with suppression of artificial disturbances and preservation of flow topology and sharp flow features:
\begin{equation}
 \mathcal L
 =\mathcal L_{\mathrm d}
 +\lambda_{\mathrm f}\mathcal L_{\mathrm f}
 +\lambda_{\mathrm s}\mathcal L_{\mathrm s}
 +\lambda_{\mathrm j}\mathcal L_{\mathrm j}.
 \label{eq:loss}
\end{equation}
Let $z_{ic}$ and $\widehat z_{ic}$ denote the normalized target and prediction for channel $c$ at cell $i$. For an internal face $e$ connecting owner cell $o(e)$ and neighbor cell $n(e)$, define the cell--to--cell difference as
$\Delta_e \widehat z_c= \widehat z_{o(e),c}- \widehat z_{n(e),c}$. The four loss terms are
\begin{equation}
\begin{aligned}
 \mathcal L_{\mathrm d} &= \left\langle w^{\mathrm d}_{ic} \bigl(\widehat z_{ic}-z_{ic}\bigr)^2 \right\rangle_{\mathcal C,c}, \\
 \mathcal L_{\mathrm f} &= \left\langle w^{\mathrm f}_{i} \bigl|\widehat z_{ic}\bigr| \right\rangle_{\mathcal C,c}, \\
 \mathcal L_{\mathrm s} &= \left\langle w^{\mathrm s}_{e} \bigl(\Delta_e\widehat z_c\bigr)^2 \right\rangle_{\mathcal F_{\mathrm s},c}, \\
 \mathcal L_{\mathrm j} &= \left\langle w^{\mathrm j}_{ec} \left( \frac{\Delta_e\widehat z_c-\Delta_e z_c} {|\Delta_e z_c|+\varepsilon_{\mathrm j}} \right)^2 \right\rangle_{\mathcal F_{\mathrm j},c}.
\end{aligned}
\label{eq:loss_terms}
\end{equation}
Here $\mathcal C$, $\mathcal F_{\mathrm s}$, and $\mathcal F_{\mathrm j}$ denote the collections of sampled cells, smoothness faces, and jump-matching faces, respectively. Angle brackets $\langle\cdot\rangle$ denote averages over the indicated samples and output channels.
The nonnegative weights incorporate the applicable regional and channel weighting, and, for jump matching, importance--sampling corrections. The positive floor $\varepsilon_{\mathrm j}$ regularizes the relative jump mismatch.

The data term $\mathcal L_{\mathrm d}$ penalizes squared prediction errors, with additional weight near walls to improve boundary--layer prediction. The freestream term $\mathcal L_{\mathrm f}$ penalizes the absolute predicted perturbations away from walls where the reference mean--flow field is close to freestream. This penalty is weakened in wakes and other disturbed regions to preserve physical departures from freestream, while suppressing small artificial perturbations that can accumulate to large training loss value over a large fraction of the mesh.

The smoothness term $\mathcal L_{\mathrm s}$ suppresses spurious cell--to--cell oscillations in smooth regions away from walls. Its weight decreases near walls and where the target varies sharply, limiting artificial smoothing of boundary layers, shocks, and shear layers. The Sobolev-type jump-matching term $\mathcal L_{\mathrm j}$ complements this penalty by matching predicted and target cell-to-cell differences between neighboring cells, with additional emphasis on pressure and temperature. Its relative scaling limits the dominance of large target jumps, while the positive floor controls amplification where target differences approach zero. \ref{app:objective} specifies further details related to the weights, masks, and scaling.

The loss balance is motivated by the differing rates at which the CFD solver removes initialization errors. Persistent errors in the pressure and velocity distributions, shock placement, and separated-flow structure can prolong convergence even after local disturbances have decayed. The objective therefore emphasizes field fidelity and feature matching while limiting the influence of freestream and smoothness regularization.

To maintain this balance during training, the auxiliary coefficients $\lambda_{\mathrm f}$, $\lambda_{\mathrm s}$, and $\lambda_{\mathrm j}$ are adjusted using parameter--gradient norms, following the motivation of gradient--based balancing~\citep{ChenEtAl2018GradNorm,WangEtAl2021Gradient}. Each weighted term's gradient norm is expressed as a fraction of the sum of these norms, and the auxiliary coefficients are updated toward prescribed shares. These shares encode the intended training priorities; they do not directly measure CFD error-decay rates. Measurements are smoothed across epochs, and updates are damped and bounded. The data coefficient remains unity, and all coefficients are held fixed during the final training stage. No CFD residual or force-coefficient loss is used.

\subsection{Sampled training and full-mesh inference}
\label{sec:training}
Based on the idea employed for Transolver--3~\citep{ZhouEtAl2026Transolver3}, the model is trained on repeatedly sampled subsets of the native mesh. This reduces computation and the memory needed to store intermediate network values for backpropagation, making training on large industrial meshes more tractable.

Each sampled subset combines three sample groups: 170,000 uniformly drawn \emph{cells} throughout the flow field for the data and freestream losses; 24,000 uniformly drawn \emph{internal faces} for smoothness; and 12,000 \emph{internal faces} for Sobolev-type jump matching. The last group comprises 6,000 uniformly drawn faces and 6,000 drawn preferentially according to the magnitude of target differences between neighboring cells. The uniform groups provide broad coverage for field accuracy and noise suppression, while preferential sampling increases exposure to sharp variations associated with shocks, shear layers, and wake boundaries. 

Sampling allows repeated selection of the same cell or face within a training step. The sampled cells and both cells adjacent to each sampled internal face are combined, with duplicates removed, into one network input containing at most 242,000 distinct cells. Each unique cell is evaluated once, but repeated samples retain their original frequency in the loss averages. Each training case supplies five newly sampled updates per epoch; an epoch therefore does not constitute a complete sweep over the mesh. Attention aggregates information from the sampled cells, so its slice states can differ from those obtained using the full mesh. Repeated resampling across epochs provides broad mesh coverage, with many cells visited multiple times during training.

Inference evaluates all cells of each mesh in chunks of at most 500,000 cells. Each Transolver block first combines information from all chunks into global slice states, then updates each chunk using these shared states. This preserves communication across the full mesh and matches unchunked evaluation up to floating--point differences. For a fixed architecture, computational work grows linearly with mesh cell count. Chunking reduces temporary memory use.
~\ref{app:architecture} and ~\ref{app:objective} provide further implementation details.

\subsection{Field reconstruction and solver integration}
\label{sec:writeback}
Reconstruction reverses Eq.~\eqref{eq:targets}, restores turbulence variables from logarithmic form, and clips their internal--cell values to the specified minima. Solver--specific interfaces handle boundary conditions, derived quantities, and restart metadata.~\ref{app:protocol} details the additional integration requirements for HiSA solver~\citep{HeynsEtAl2014}. 

The reconstructed fields initialize the RANS solver, completing the prediction--to--solver sequence shown in Fig.~\ref{fig:schematic}(A). Evaluation measures the subsequent solver work and assesses agreement between converged warm-- and cold--start solutions under matched numerical settings. The next section presents the datasets and convergence criteria, with supporting details of solver initialization and assessment provided in~\ref{app:protocol}.

\section{Computational Experiments and Assessment}
\label{sec:experiments}
\subsection{Datasets and Experimental Design}
\label{sec:datasets}
\begin{figure}[t]
\centering
\begin{subfigure}{0.32\textwidth}
\includegraphics[width=\linewidth]{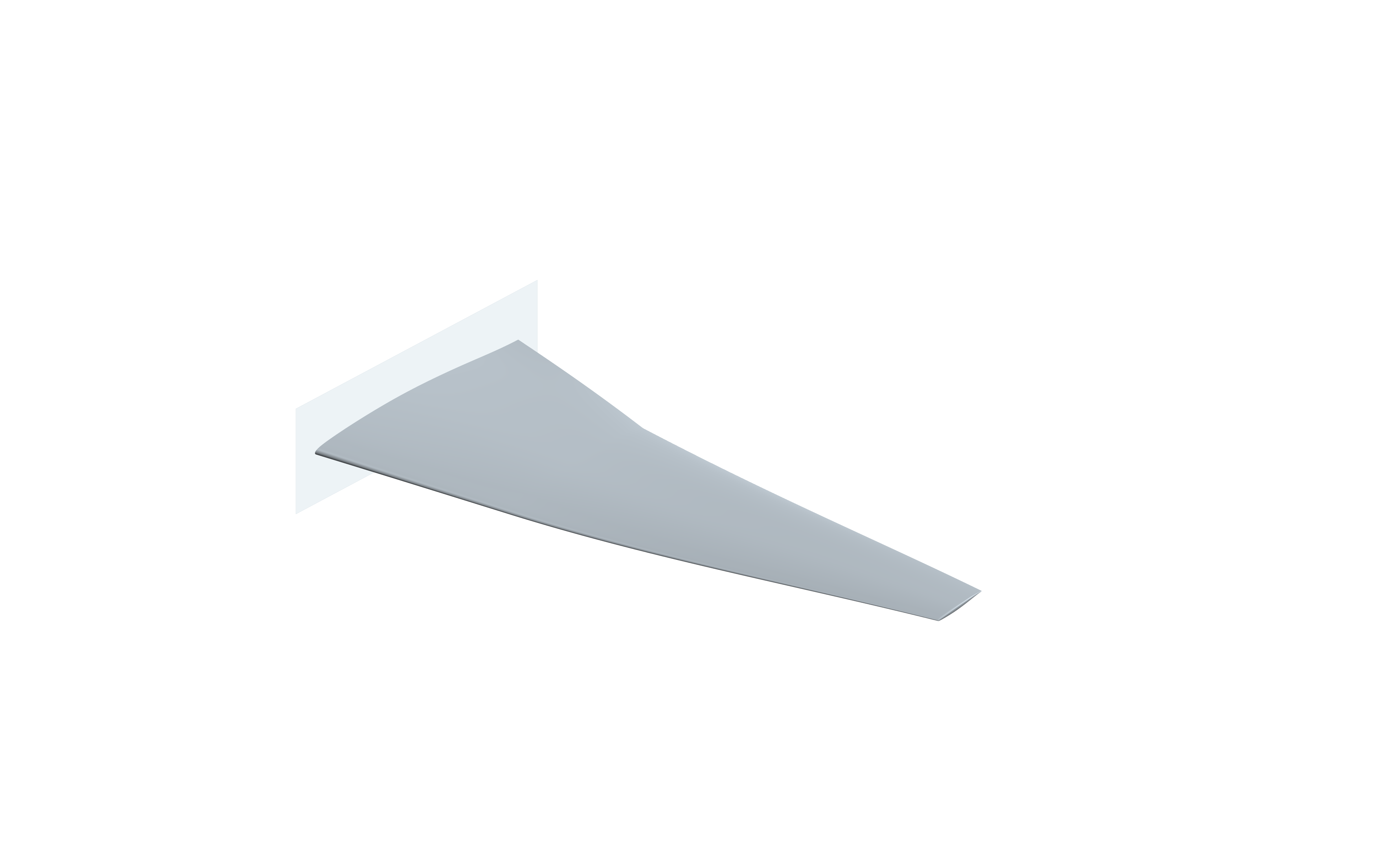}
\caption{SuperWing Geometry}
\end{subfigure}
\begin{subfigure}{0.33\textwidth}
\includegraphics[width=\linewidth]{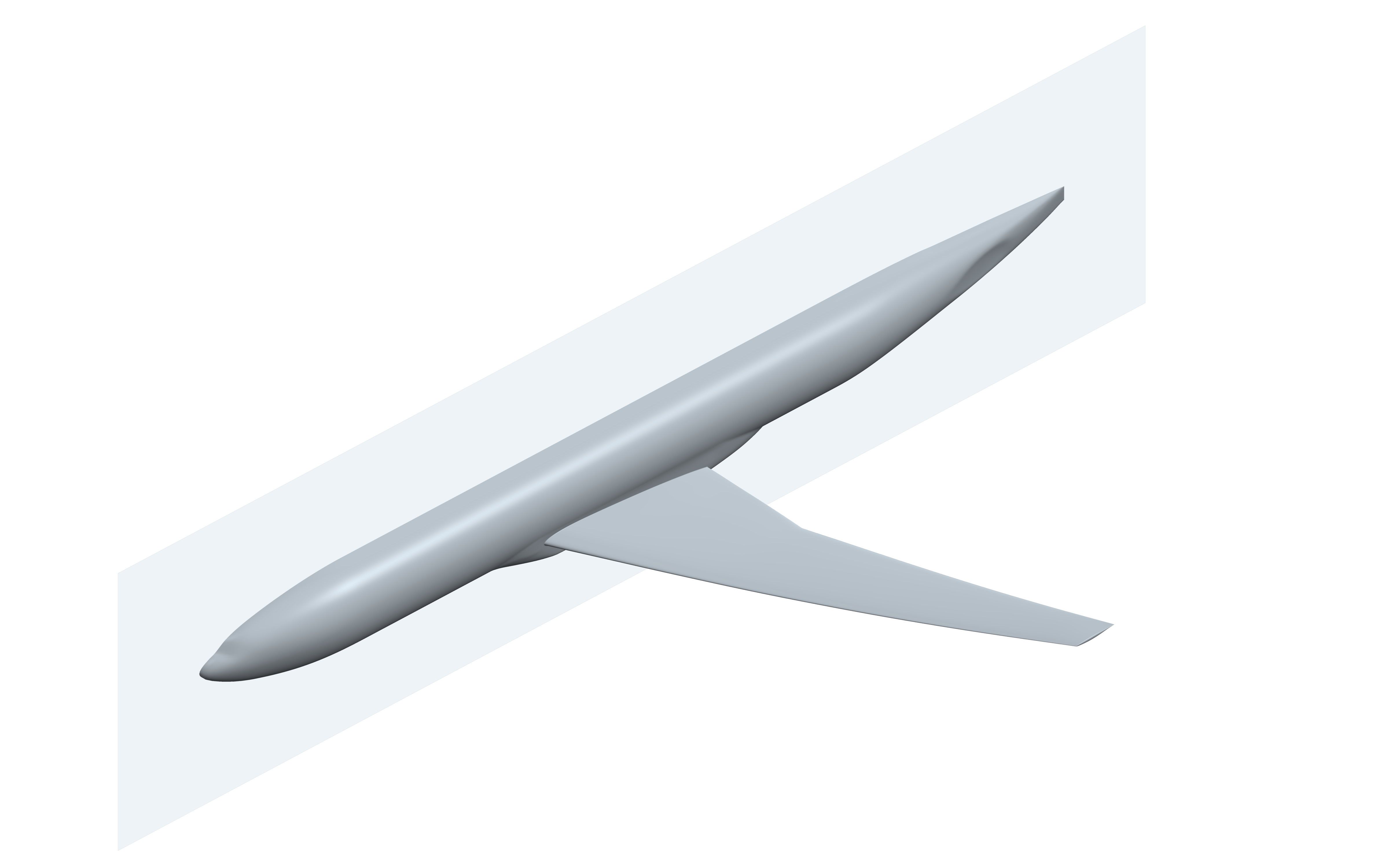}
\caption{NASA CRM Model}
\end{subfigure}
\begin{subfigure}{0.33\textwidth}
\includegraphics[width=\linewidth]{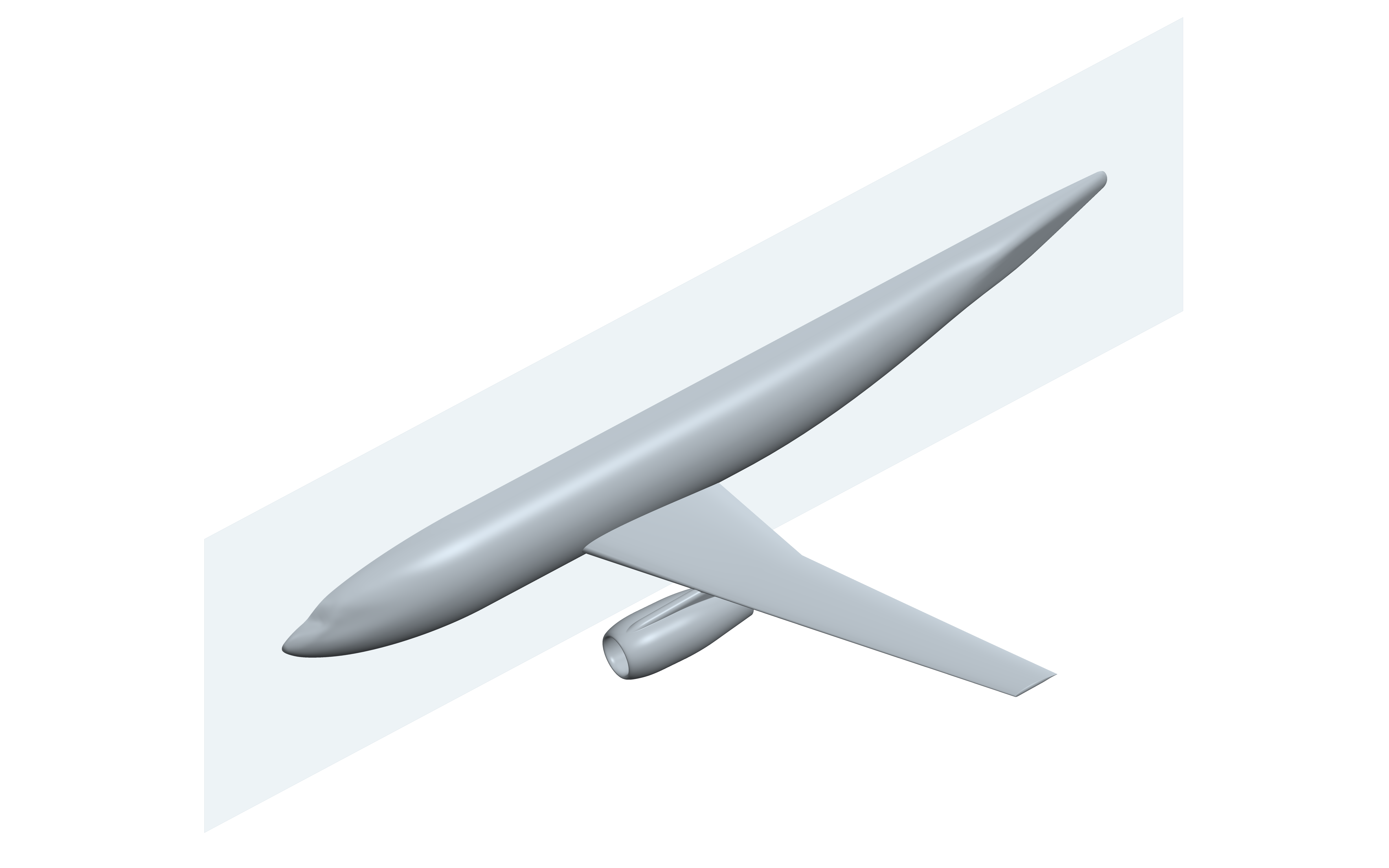}
\caption{DLR F6 Model}
\end{subfigure}
\caption{Aerodynamic configurations used to assess learned warm starts. Half--model geometries are shown, with shaded planes indicating the symmetry boundaries of the computational domains.}\label{fig:geometries}
\end{figure}

The computational experiments assess learned initialization using three complementary aerodynamic datasets: SuperWing, the NASA Common Research Model (CRM), and DLR-F6. Representative geometries are shown in Fig.~\ref{fig:geometries}. SuperWing provides geometric variation through a family of swept wings~\citep{YangEtAl2025SuperWing}. The CRM wing--body configuration provides a fixed geometry over a broader operating envelope~\citep{VassbergEtAl2008,LevyEtAl2014}, while the DLR--F6 wing--body--nacelle--pylon configuration extends the assessment to a different aircraft geometry and turbulence closure~\citep{LaflinEtAl2005}. Unless otherwise stated, all training fields are generated using the HiSA solver in OpenFOAM, with the Spalart--Allmaras turbulence model for SuperWing and CRM and the $k$--$\omega$ SST model for DLR--F6. The geometries and mesh--generation procedures are adopted from the respective sources cited above, and the flow solutions are recomputed for the present study.

Tables~\ref{tab:superwing_dataset}--\ref{tab:f6_dataset} summarize the datasets. SuperWing emphasizes geometry--dependent swept--wing shocks, their interaction with the boundary layer, and localized separation. The CRM cases span subcritical and transonic conditions, capturing shock onset and strengthening, changes in pressure recovery, and side--of--body separation at more demanding operating points. DLR--F6 introduces additional wing--body--nacelle--pylon interference, with junction separation and shock--boundary--layer interactions as distinguishing features. Together, these datasets assess warm starts across geometric variation, challenging flow regime, and increasingly complex component interactions.
\begin{table}[t]
\centering
\small
\caption{SuperWing dataset used in this study. Geometric ranges cover
all retained wings.}
\label{tab:superwing_dataset}
\begin{tabular}{@{}p{0.57\linewidth}p{0.36\linewidth}@{}}
\toprule
Parameter & Range or value \\
\midrule
Leading-edge sweep angle               & $25.10^\circ$--$39.43^\circ$ \\
Trapezoidal aspect ratio               & $8.05$--$10.87$ \\
Trapezoidal taper ratio                & $0.151$--$0.398$ \\
Kink location, fraction of semispan    & $0.360$--$0.417$ \\
Root thickness-to-chord ratio          & $0.1409$--$0.1682$ \\
Tip thickness-to-chord ratio           & $0.0765$--$0.1024$ \\
Leading-edge dihedral parameter at tip & $4.04^\circ$--$9.40^\circ$ \\
Tip washout relative to the centreline & $8.76^\circ$--$13.42^\circ$ \\
\midrule
Freestream Mach number, $M_\infty$ & $0.83$ \\
Angle of attack, $\alpha$          & $\{6^\circ,\,7^\circ,\,8^\circ\}$ \\
Reynolds number, $Re_{L_{\mathrm{ref}}}$ & $2\times10^7$ \\
Distinct geometries & $32$ \\
Coarse mesh, L1 & $0.44$\,M cells \\
Fine mesh, L2 & $3.53$\,M cells \\
Total flow cases & $96$ per mesh level \\
\bottomrule
\end{tabular}
\end{table}

\begin{table}[t]
\centering
\small
\caption{CRM operating conditions and case distribution. Ranges denote the extrema of the sampled conditions; the out--of--distribution (OOD) conditions are listed individually. The wing--body configuration uses two levels of DPW--5 meshes.}
\label{tab:crm_dataset}
\begin{tabular}{@{}lcccc@{}}
\toprule
Case group & $M_\infty$ & $\alpha\ (^\circ)$
           & $Re_{L_{\mathrm{ref}}}/10^6$ & Cases \\
\midrule
Nominal design & $0.60$--$0.87$ & $-1.50$--$4.00$ & $1.00$--$7.00$ & $103$ \\
Off-grid & $0.617$--$0.855$ & $-0.48$--$3.78$ & $1.22$--$6.11$ & $14$ \\
\midrule
OOD: lower Mach number & $0.55$ & $2.50$ & $5.00$ & $1$ \\
OOD: higher Mach number & $0.90$ & $2.50$ & $5.00$ & $1$ \\
OOD: higher incidence & $0.60$ & $5.50$ & $5.00$ & $1$ \\
OOD: higher Reynolds number & $0.87$ & $1.25$ & $9.00$ & $1$ \\
\midrule
Coarse mesh, L2 & & & & $2.16$\,M cells \\
Fine mesh, L3 & & & & $5.11$\,M cells \\
Total flow cases & & & & $121$ per mesh level \\
\bottomrule
\end{tabular}
\end{table}

\begin{table}[t]
\centering
\small
\caption{DLR--F6 operating conditions and case distribution. Ranges denote the extrema of the sampled conditions; the out--of--distribution (OOD) conditions are listed individually. The wing--body--nacelle--pylon configuration uses a DPW--2 mesh of $4.79$ M cells.}
\label{tab:f6_dataset}
\begin{tabular}{@{}lcccc@{}}
\toprule
Case group & $M_\infty$ & $\alpha\ (^\circ)$
           & $Re_{L_{\mathrm{ref}}}/10^6$ & Cases \\
\midrule
Nominal design
    & $0.60$--$0.80$ & $-2.00$--$3.50$ & $1.50$--$4.50$ & $37$ \\
Off-grid
    & $0.609$--$0.790$ & $-1.75$--$3.05$ & $1.60$--$4.19$ & $8$ \\
\midrule
OOD: higher Mach number
    & $0.82$ & $1.00$ & $3.00$ & $1$ \\
OOD: higher incidence
    & $0.75$ & $4.50$ & $3.00$ & $1$ \\
\midrule
Total flow cases & & & & $47$ \\
\bottomrule
\end{tabular}
\end{table}

Training subsets are drawn from the nominal case pools using seeded randomization balanced across operating--condition levels. For SuperWing, training uses two angles per selected geometry from 26 in-range wings; Test comprises the remaining cases of these wings, including withheld angles on trained geometries and cases on untrained geometries. The six out--of--distribution (OOD) wings provide 18 evaluation-only cases. For CRM and DLR--F6, training sets are drawn using the same process while Test comprises the remaining nominal conditions, while off-grid and OOD conditions are reserved for evaluation only. SuperWing and CRM use coarse-mesh training, with the trained models also evaluated on finer meshes. Construction and partitioning details are given in \ref{app:datasets}.

Each model is trained independently using the common architecture and objective structure of Section~\ref{sec:formulation}, with closure--specific output channels. The optimization steps per training case is held fixed within each dataset; consequently, larger training sets also receive more parameter updates. The resulting learning curves assess the benefit of additional cases under this training policy. Final-epoch weights are used for subsequent inference. \ref{app:architecture} specifies the training settings.

\subsection{Assessment Protocol and Comparison Baselines}
\label{sec:protocol}
Each warm run is paired with a cold run initialized from the uniform freestream state, the solver's standard initialization. The paired runs use identical meshes, boundary conditions, discretization, pseudo-time controls and core counts; only their initial internal fields differ (\ref{app:restart}). Warm runs use fixed budgets of 5,000--15,000 iterations, with convergence assessed retrospectively.

Convergence is assessed using the integrated lift, drag and pitching-moment coefficients ($C_L$, $C_D$ \& $C_M$), with both (cold and warm) starts evaluated against the same cold--run reference. At the primary tolerance of $\epsilon=1\%$, each assessed coefficient must remain within its tolerance band for the remainder of the run. Its settling iteration is recorded 100 iterations after the final band entry. Reference values construction and convergence measures are detailed in \ref{app:convergence}. A very few of the warm runs that fail this criterion are excluded from speed--up averages and are analyzed in Section~\ref{sec:robustness}. Sensitivity to tolerances of $\epsilon=0.5\%$ and $\epsilon=2\%$ is also reported in \ref{app:tolerance}.

Per--case acceleration is measured by $s=N_{\mathrm{cold}}/N_{\mathrm{warm}}$ and summarized for multiple flow cases by the geometric mean
\begin{equation}
  \bar{s}_\mathrm{g}
  =\exp\Bigg(\frac{1}{n}\sum_{a=1}^{n}\ln s_a\Bigg),
  \label{eq:geomean}
\end{equation}
which balances reciprocal gains and losses and is less sensitive than the arithmetic mean to exceptionally large speed-ups \citep{FlemingWallace1986}. Arithmetic mean in this case would provide a slightly higher value of mean speed--up. Aggregate solver costs are compared through the pooled ratio
\begin{equation}
  S=\frac{\sum_a C_{\mathrm{cold},a}}{\sum_a C_{\mathrm{warm},a}},
  \label{eq:pooled}
\end{equation}
where $C$ is measured in core-hours. Full campaign comparisons additionally account for one-off data-generation and model-training costs (Section~\ref{sec:cost}).

Within each dataset, flow cases are grouped by role, where applicable:
\begin{itemize}[nosep]
\item \emph{Train}: training cases, with warm runs used for diagnostics.
\item \emph{Test}: remaining nominal cases.
\item \emph{Off--grid}: additional conditions drawn by Latin hypercube sampling within the nominal training range.
\item \emph{OOD}: conditions or geometries outside the training range.
\end{itemize}
As the nested training sets grow, Test shrinks, while Off-grid and OOD remain fixed.


\subsection{SuperWing: Geometric Generalization and Cross--Mesh Performance}
\label{sec:superwing}
\begin{figure}[t]
\centering
\includegraphics[width=0.42\textwidth]{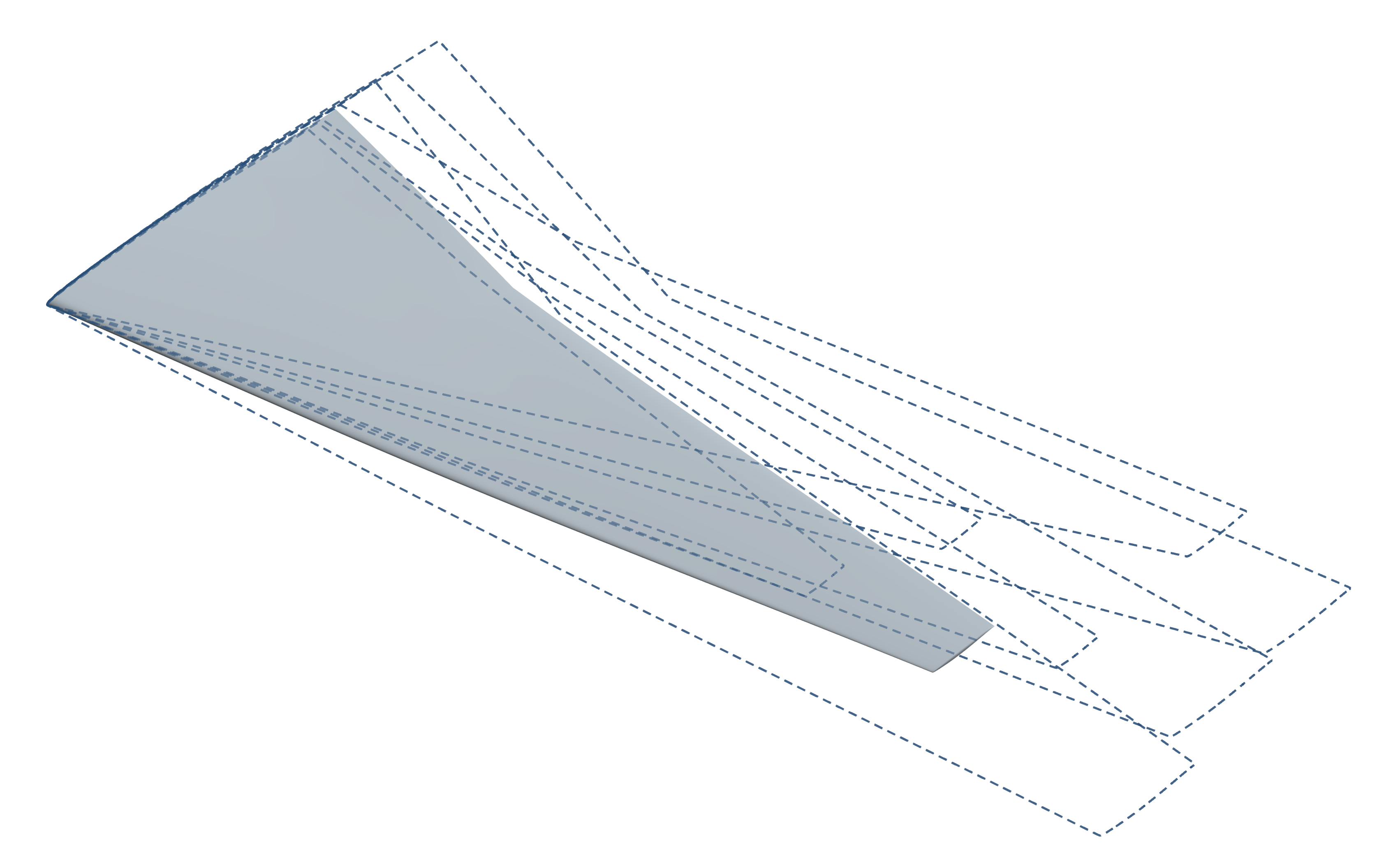}
\caption{SuperWing planform variation: representative wings overlaid on a shaded reference wing and aligned at the root leading edge.}
\label{fig:sw_family}
\end{figure}

SuperWing varies geometry and incidence at fixed Mach and Reynolds numbers: 32 distinct wing shapes at $\alpha=6^\circ$, $7^\circ$ and $8^\circ$ give 96 cases per mesh (Table~\ref{tab:superwing_dataset}). At fixed incidence, the range of $C_L$ across wing geometries is 0.27--0.33, comparable to the median range of 0.23 obtained by varying incidence across the three angles for each wing. Hence flow solution varies significantly along these two axes. Training subsets select 4--16 of the 26 in-range wings, with two incidences each ($N=8$--32; \ref{app:partitions}). The withheld incidence tests interpolation ($7^\circ$) or extrapolation ($6^\circ$, $8^\circ$) on trained geometries; untrained wings test geometric generalization. Six wings outside the pool’s geometric parameter range form the 18-case OOD set, differing chiefly in taper ratio: 0.15--0.20 for five wings, compared with 0.23--0.40 in the pool. Figure~\ref{fig:sw_family} illustrates the planform variation.

\begin{figure}[t]
\centering
\includegraphics[trim={0 3mm 0 14mm}, clip, width=0.63\textwidth]{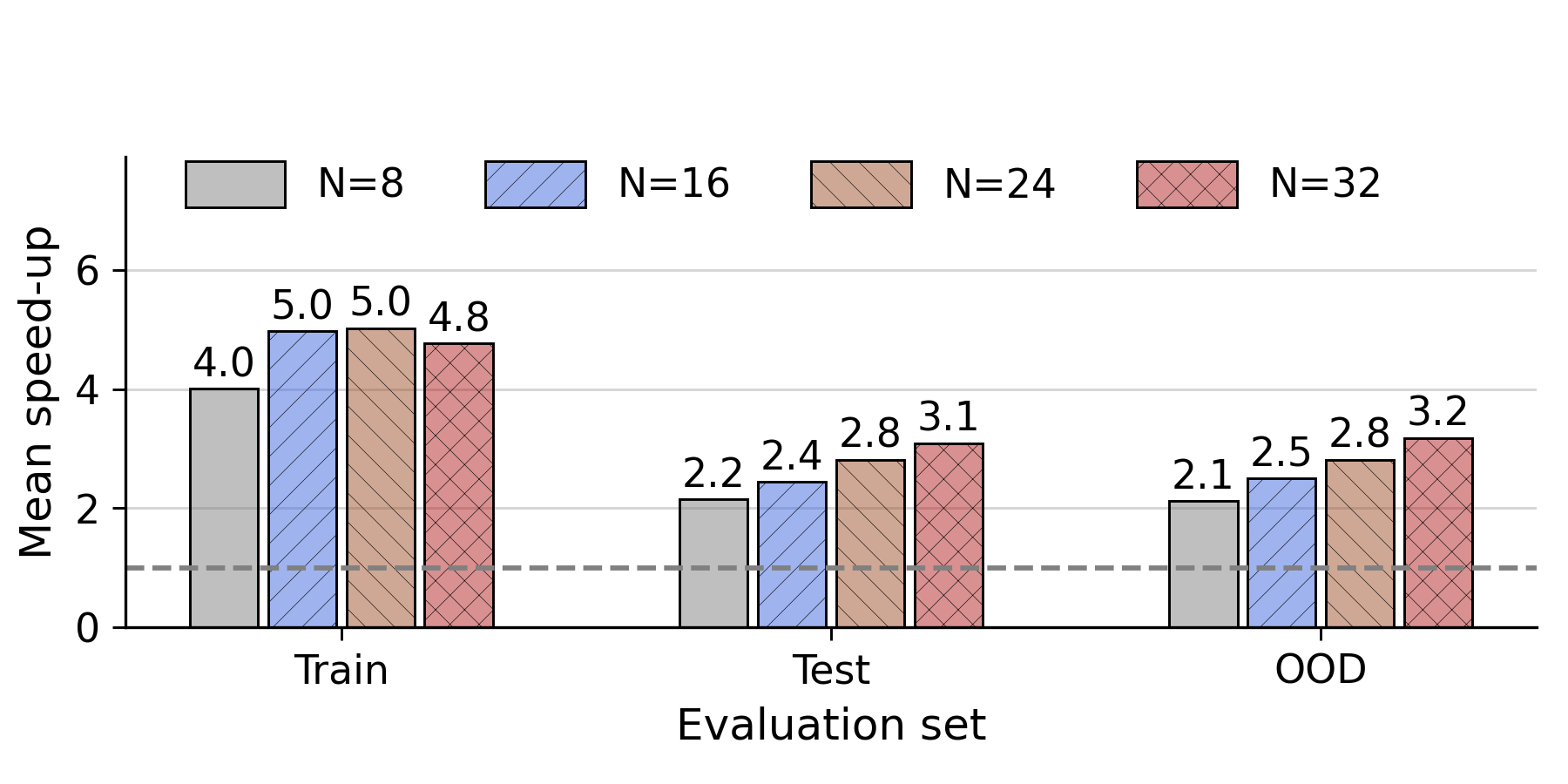}
\caption{SuperWing on L1: geometric--mean speed--up at the 1\% criterion versus training--set size $N$. Train: training cases; Test: remaining in--range cases; OOD: 18 cases on six out--of--range wings. The dashed line marks unit speed--up.}
\label{fig:sw_speedup}
\end{figure}

All 96 warm runs converge at the 1\% criterion for every $N$ (Fig.~\ref{fig:sw_speedup}). Train speed-up plateaus early, while Test and OOD speed-ups continue to improve with training-set size. The fixed OOD population provides the clearest evidence of this improvement, since Test shrinks as training expands. Results also demonstrate that generalization across geometry is more demanding than across incidence: withheld incidences on trained wings approach Train performance, whereas untrained in-range wings achieve smaller gains. Across datasets, untrained wings have roughly four times the relative perturbation error of withheld CRM and DLR-F6 conditions, yet achieve acceleration of similar magnitude (Eq.~\eqref{eq:field-error}; Sections~\ref{sec:crm} and~\ref{sec:f6}). Pointwise accuracy alone therefore does not determine warm-start effectiveness.

Cross--mesh inference demonstrates resolution invariance: the same trained model operates at a different mesh resolution without retraining. This is enabled by shared cell--wise transformations and global slice aggregation, whose learned parameters are independent of cell count and mesh connectivity. The $N=24$ model trained on L1~(0.44\,M cells) is applied to L2~(3.53\,M cells), which doubles the resolution in each direction. Features are recomputed on L2 while retaining the L1 normalization statistics and physical Fourier wavelengths (Section~\ref{sec:training}); no fine--mesh training data, solution interpolation or target--condition coarse solve is required.

\begin{figure}[t]
\centering
\includegraphics[trim={0 3mm 0 14mm}, clip, width=0.52\textwidth]{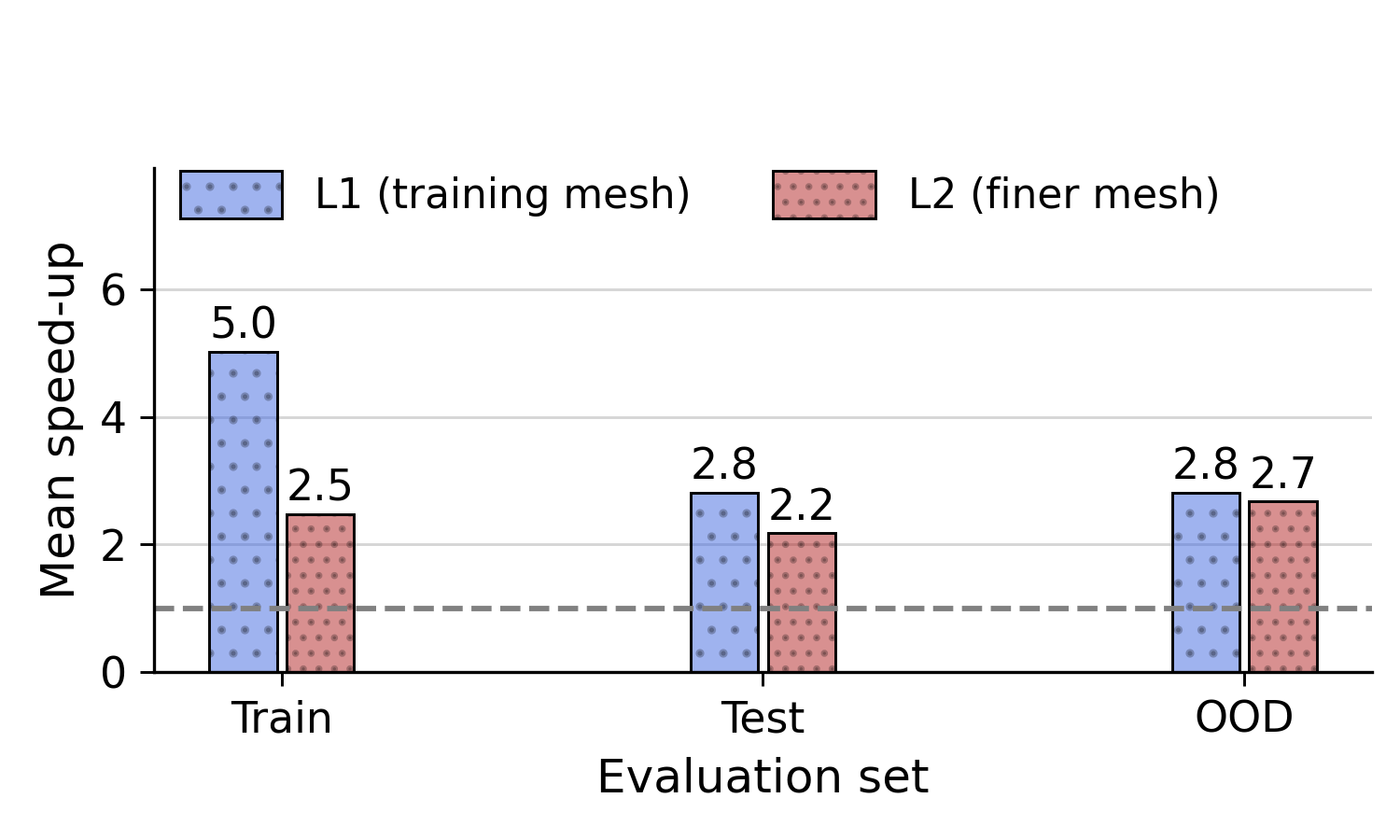}
\caption{SuperWing cross--mesh inference: geometric--mean speed--up at the 1\% criterion for the $N=24$ model trained on L1 and evaluated on L1 and L2, relative to each mesh's cold runs.}
\label{fig:sw_speedup_xres}
\end{figure}

All L2 warm runs converge (Fig.~\ref{fig:sw_speedup_xres}). Geometric mean speed--ups become more similar across categories: Train cases lose much of their coarse--mesh advantage, while OOD wings retain nearly all their acceleration. This pattern is consistent with the fine--mesh solver correcting inherited coarse--mesh discretization error, even for trained conditions: L1 drag exceeds L2 drag by a median of 20\%, mainly through pressure drag. Near--wall errors associated with first--cell distances below the training range, however, decay rather swiftly in the cases examined, suggesting that these local discrepancies do not govern the longer settling process. Resolution invariance enables cross--mesh deployment, while mesh--dependent solver corrections alter the achieved speed--up without eliminating the acceleration benefit.

Importantly, these results demonstrate acceleration at two levels in cross--mesh deployment: L1 training data generation costs approximately 21 times less than the L2 solutions, yet the resulting model accelerates L2 runs by $2.3\times$ overall, with gains extending to untrained and OOD geometries. This suggests potential use in aerodynamic shape optimization, where a model trained on coarse solutions could initialize fine--mesh evaluations of modified designs, for example when minimizing drag at prescribed lift. Previous--design restarts are effective for small shape changes \citep{WangEtAl2020Remapping}; learned initialization requires no previous--design solution and could support larger shape changes and simultaneous CFD evaluations of multiple candidate designs. Under the campaign--cost assessments of Section~\ref{sec:cost}, the initial investment is recovered after about four fine--mesh evaluations.


\subsection{NASA CRM: Operating-Envelope Generalization and Cross-Mesh Transfer}
\label{sec:crm}
The CRM dataset varies operating conditions on a fixed wing--body geometry, from subcritical flow to shock--induced separation (Table~\ref{tab:crm_dataset}). Nominal conditions form a sparse Ma--$\alpha$--Re grid; off--grid cases test joint interpolation, while four OOD cases each extend beyond one parameter's nominal range. Nested training subsets balance Reynolds--number representation within Mach--$\alpha$ groups (\ref{app:partitions}); the remaining nominal cases form Test. Models are trained on L2 mesh ($2.16$\,M), with the $N=24$ model also evaluated on L3 mesh ($5.11$\,M).

\begin{figure}[t]
\centering
\includegraphics[trim={0 3mm 0 14mm}, clip, width=0.67\textwidth]{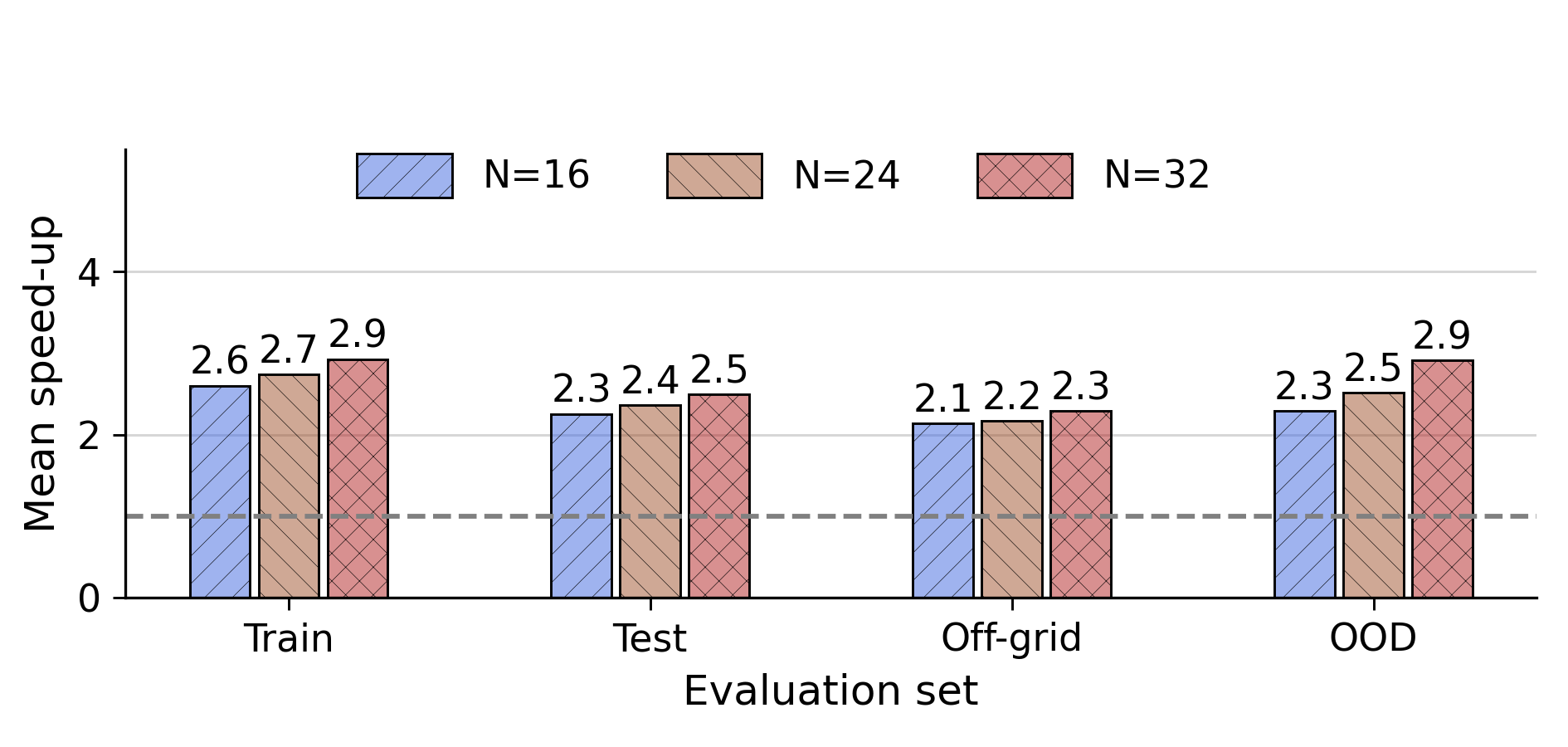}
\caption{CRM on the training mesh (DPW-5 L2, 2.16\,M cells): geometric--mean speed--up at the 1\% convergence criterion for different training--set sizes (N). Test: held--out nominal conditions; Off-grid: 14 conditions sampled using Latin hypercube sampling; OOD: four conditions outside the nominal range.}\label{fig:crm_speedup}
\end{figure}

Of 121 warm runs, 119 converge for every $N$; two consistently differ from the cold reference and are discussed in Section~\ref{sec:robustness}. As shown in Fig.~\ref{fig:crm_speedup}, Test, Off--grid and OOD speed--ups improve with training--set size. The smaller Train--Test gap than for SuperWing indicates effective generalization across operating conditions on a fixed geometry. Off--grid cases yield generally smaller but consistent gains despite varying all three parameters simultaneously, while the improving OOD trend is based on only four cases.

\begin{figure}[t]
\centering
\includegraphics[trim={0 3mm 0 14mm}, clip, width=0.54\textwidth]{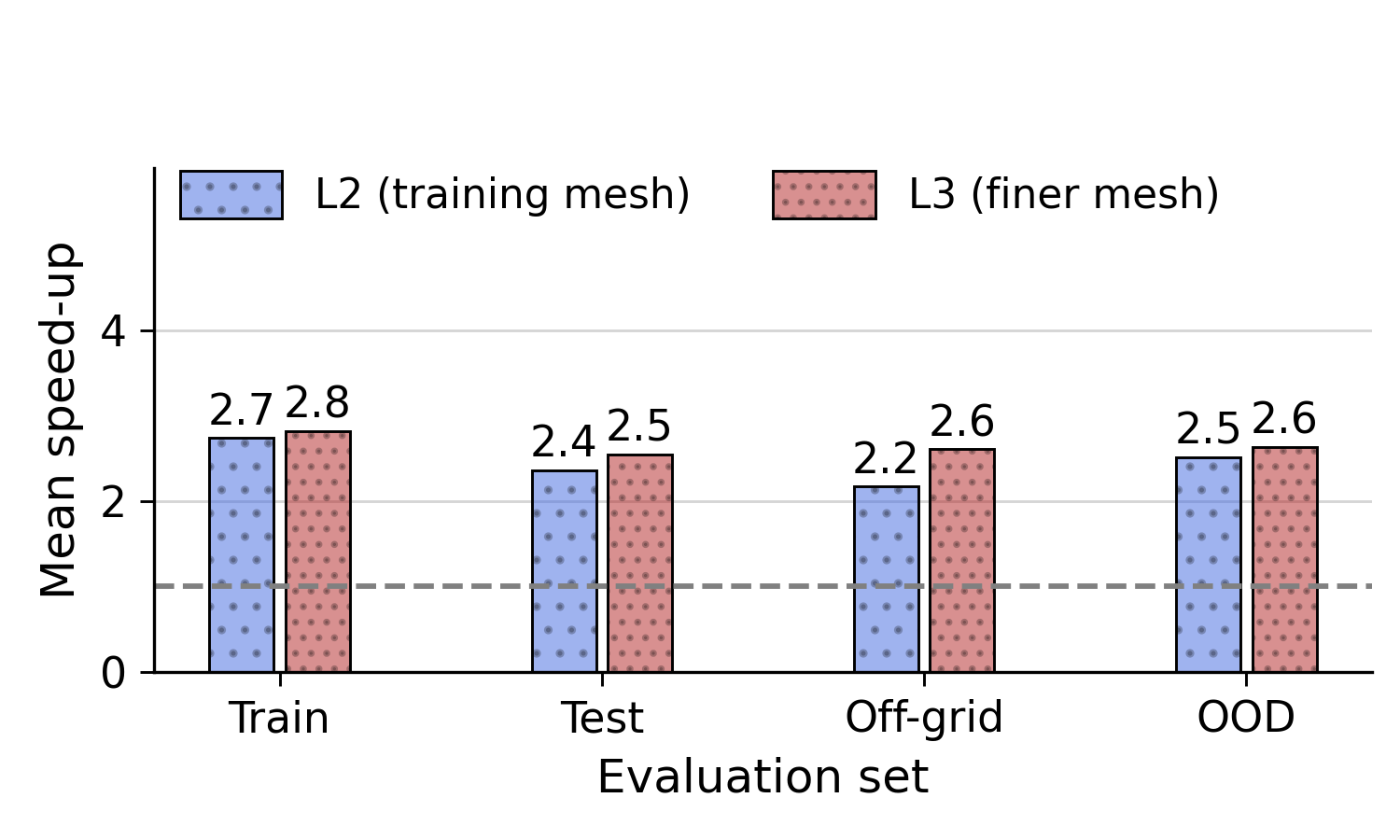}
\caption{CRM cross--mesh inference: geometric--mean speed--up at the 1\% convergence criterion for the $N=24$ model trained on the DPW-5 L2 mesh and evaluated on both L2 and the finer L3 mesh (5.11\,M cells), relative to the corresponding cold--start runs on each mesh.}\label{fig:crm_speedup_xres}
\end{figure}

Applying the $N=24$ model trained on L2 directly to L3 increases the geometric--mean speed--up in every category (Fig.~\ref{fig:crm_speedup_xres}), as refinement increases cold--run iteration counts more than warm--run counts. Attached flows benefit, whereas cases with appreciable separation generally show lower speed--ups than on L2, consistent with the need to establish larger separated regions on L3. All converged L3 warm runs nevertheless remain faster than their cold counterparts. Unlike SuperWing, median drag and pitching--moment differences between meshes are only a few percent, suggesting a smaller discretization correction. Pitching moment still determines the settling iteration in most L3 warm runs.

Cross--mesh acceleration thus depends on the required solution correction relative to cold--start effort. For CRM, modest differences between coarse-- and fine--mesh solutions preserve effective initialization as cold--start convergence slows. L2 training also reduces data--generation cost approximately fourfold, combining cheaper training solutions with greater fine--mesh acceleration.

\subsection{DLR-F6: Component Interference and $k$--$\omega$ SST Closure}
\label{sec:f6}
The DLR--F6 dataset extends the assessment to a wing--body--nacelle--pylon configuration with $k$--$\omega$ SST closure on a single mesh (Table~\ref{tab:f6_dataset}). Its 37 nominal conditions combine a Ma--$\alpha$ grid with Reynolds--number variations at selected nodes; eight off--grid and two OOD conditions are evaluated separately. The envelope extends beyond the lift break at high Mach numbers and incidences, encompassing shock--induced wing--body separation and nacelle--pylon interference. Training subsets contain $N=8$, 16 and 24 cases.

\begin{figure}[t]
\centering
\includegraphics[trim={0 3mm 0 14mm}, clip, width=0.6\textwidth]{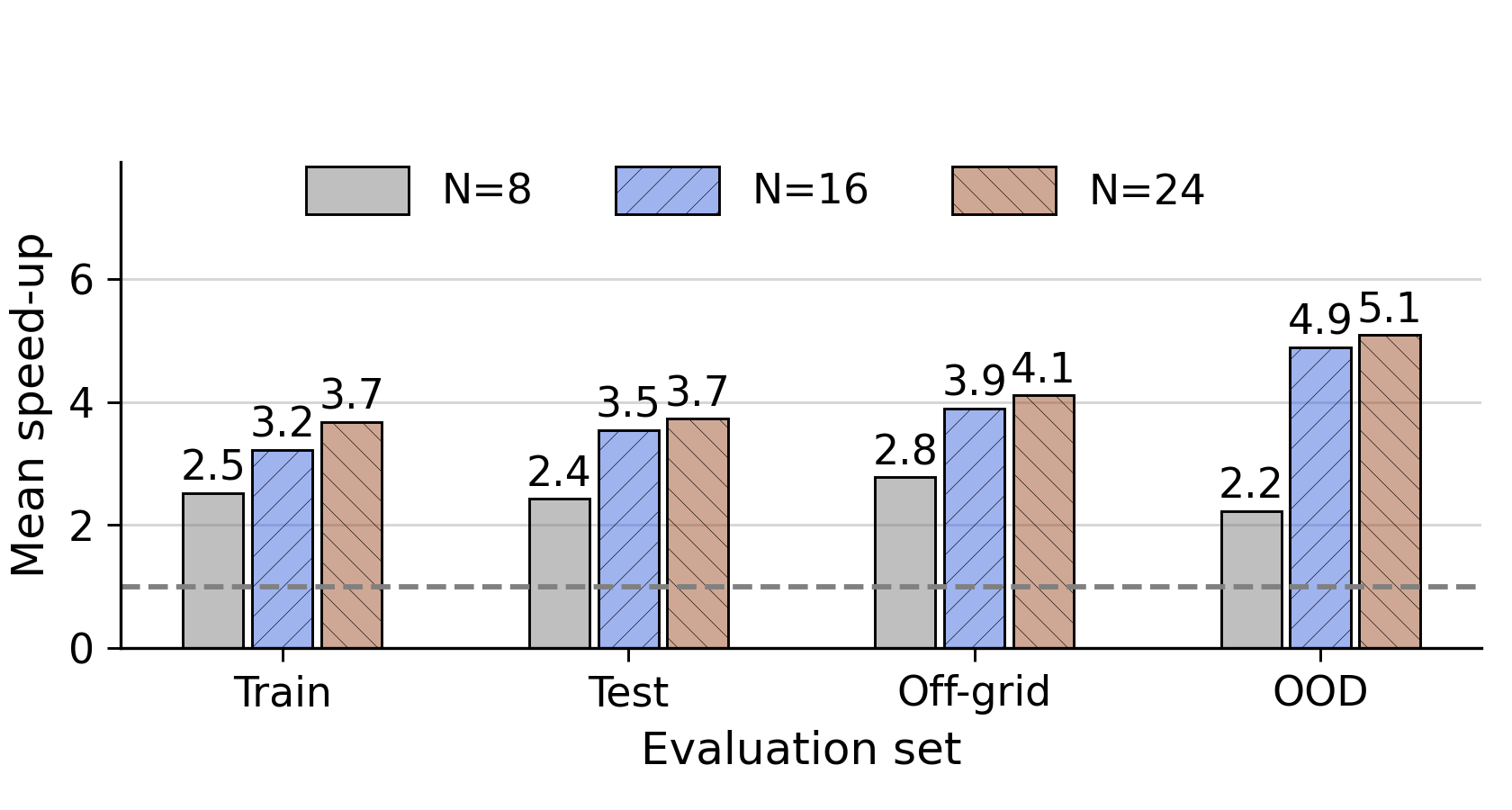}
\caption{F6 (4.79\,M cells, $k$--$\omega$ SST): geometric--mean speed--up at the 1\% convergence criterion for different training--set sizes $N$. Test: held--out nominal conditions; Off-grid: eight conditions within the nominal range; OOD: two conditions outside the nominal range.}\label{fig:f6_speedup}
\end{figure}

DLR--F6 yields the largest speed--ups of the three configurations (Fig.~\ref{fig:f6_speedup}). Test and Off--grid gains increase most strongly between $N=8$ and 16, with smaller improvements thereafter; the two OOD conditions show particularly strong acceleration. Test performance matches or exceeds Train performance for $N\ge16$, indicating effective generalization across the sampled operating envelope.

The advantage over CRM reflects shorter warm-run settling rather than slower cold runs: median cold-run counts are similar, while non-training warm runs settle in roughly half as many iterations. This advantage may partly reflect greater training coverage: at $N=24$, nearly two-thirds of the nominal DLR--F6 conditions are used for training, compared with less than one-quarter for CRM. One non--training case fails the convergence criterion for each $N$  and is discussed in Section~\ref{sec:robustness}.

\section{Discussion}
\label{sec:discussion}

\subsection{Comparison with other initialization methods}
\label{sec:baselines}
Standard aerodynamic CFD initialization includes uniform fields, potential--flow or other low--fidelity solutions, coarse--to--fine grid sequencing, and restarts from nearby conditions or previous designs. Uniform fields leave the flow structures to develop, while potential--flow estimates omit viscous effects and coarse--grid initialization requires additional target--specific computations and field transfer. Neighboring--solution restarts are a common practical alternative to cold starts, but require a suitable source and remapping when geometries or meshes differ. Sequential continuation also introduces dependencies and can influence the selected flow state near separation \citep{ZaunerEtAl2024,WangEtAl2020Remapping}. We therefore assess learned initialization against nearest--condition (NC) restarts on CRM L2, using the same 24 training solutions, predefined parameter--space pairing, freestream--scaled field transfer, and matched solver settings and convergence criteria. Details of the restart pairing procedure and freestream--scaled field transfer are given in \ref{app:nc-restart}.

\begin{figure}[t]
\centering
\includegraphics[trim={0 3mm 0 5mm}, clip, width=0.57\textwidth]{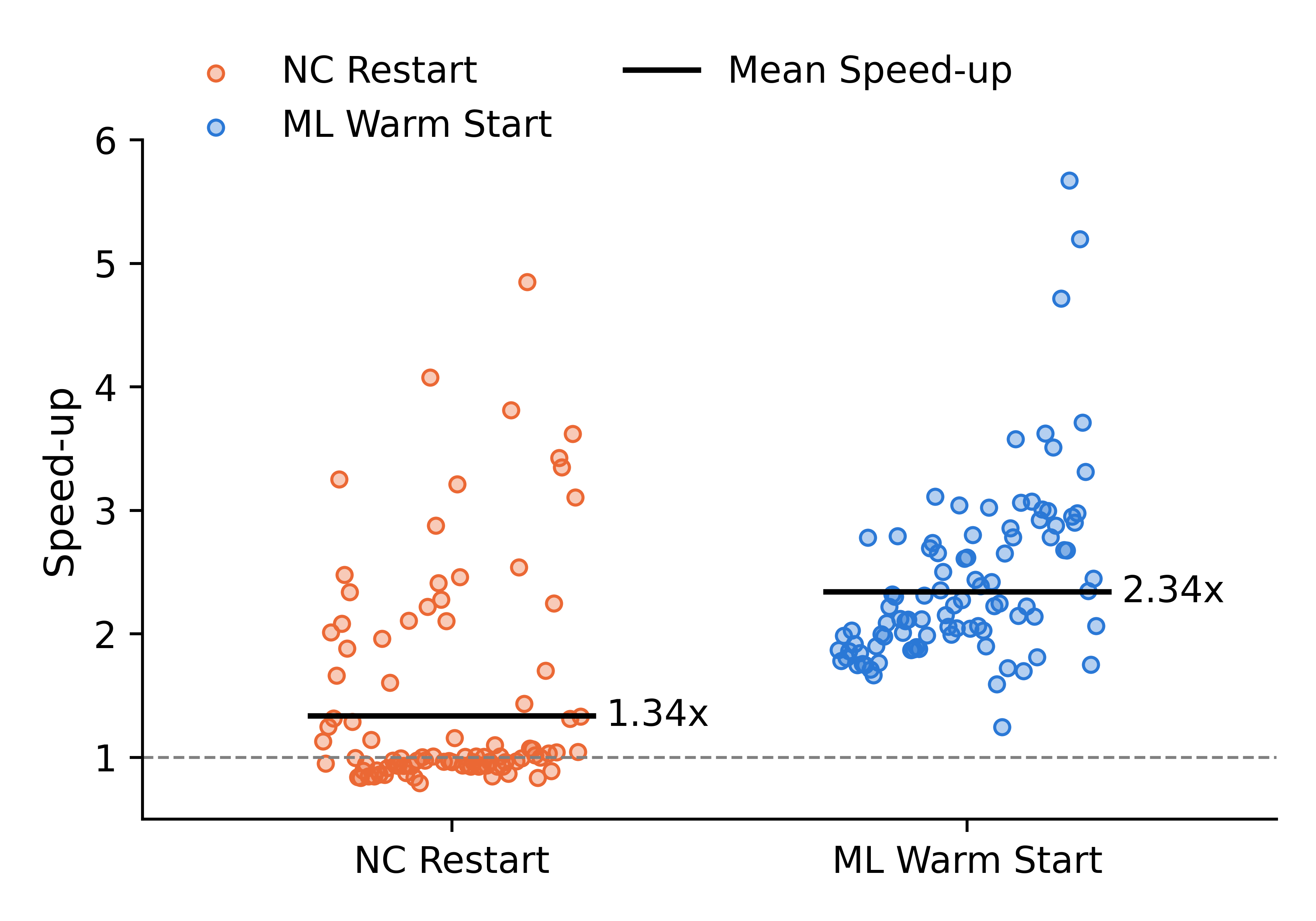}
\caption{Per--case speed--up at the 1\% convergence criterion for nearest--condition (NC) restarts and $N=24$ learned warm starts on the 95 held--out CRM cases (L2 mesh). Both methods use the same 24 training solutions. Horizontal bars indicate the geometric mean speed-ups. Dashed line marks the cold--start baseline.}\label{fig:nc_restart}
\end{figure}

Of the 97 non-training cases, 95 satisfy the common convergence criterion with both initializations. On these cases, learned initialization achieves a geometric mean speed--up of $2.34\times$, compared with $1.34\times$ for NC restarts (Fig.~\ref{fig:nc_restart}). Both methods fail to reach the cold--run reference values for the same two cases (Section~\ref{sec:robustness}). Across the 95 cases satisfying the common convergence criterion, learned initialization is faster in 74 cases and never slower than the cold start, whereas 44 NC restarts incur a slowdown compared to cold runs. Even the model trained on eight cases exceeds the aggregate performance of the 24-solution restart library. Thus, under the tested pairing rule, learning from the available solutions provides more consistent acceleration than directly reusing individual fields.

Earlier learned warm start works demonstrate acceleration but retain deployment requirements: CFDNet~\citep{ObiolsSalesEtAl2020CFDNet} uses preliminary solver iterations, super--fidelity~\citep{ZhouEtAl2025SuperFidelity} requires potential--flow inputs, and wake--extension models~\citep{FuchiEtAl2025} were assessed with supplied near--body solution data. These studies do not establish complete initialization across complex aircraft configurations and operating envelopes. MHLF provides an aircraft-scale demonstration \citep{LiuEtAl2026MHLF}, training its DLR--F6 model on 19 incidences at fixed Mach and Reynolds numbers and testing eight intermediate incidences. Our $N=16$ and $24$ models instead span all three operating parameters and are assessed across a broader evaluation set comprising withheld nominal, off-grid and OOD conditions. Geometric--mean speed--ups on the converged nominal Test cases reach $3.55$--$3.73\times$ under the common 1\% cold-reference criterion. MHLF's reported totals give a pooled iteration-count ratio of $2.31\times$ and a runtime ratio of $4.05\times$, combining initialization with asynchronous regional updates. Its workflow requires grid coarsening, regional mean-flow prediction, prolongation, and turbulence fields from neighboring conditions before CFD correction. The present model predicts mean-flow and turbulence variables directly on the target mesh while retaining the cold-run iteration schedule. This simplifies deployment and isolates the effect of initialization, although differences in meshes, solvers, convergence criteria and aggregation prevent a direct performance ranking.

\subsection{Cost Amortization and Break-Even Analysis}
\label{sec:cost}
\begin{figure}[t]
\centering
\begin{subfigure}{0.37\textwidth}
\includegraphics[trim={0 3mm 0 0}, clip, width=0.98\linewidth]{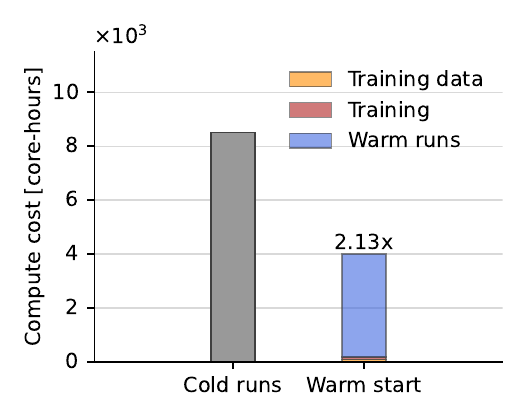}
\caption{SuperWing effective speed--up}
\end{subfigure} \qquad
\begin{subfigure}{0.37\textwidth}
\includegraphics[trim={0 3mm 0 0}, clip, width=0.98\linewidth]{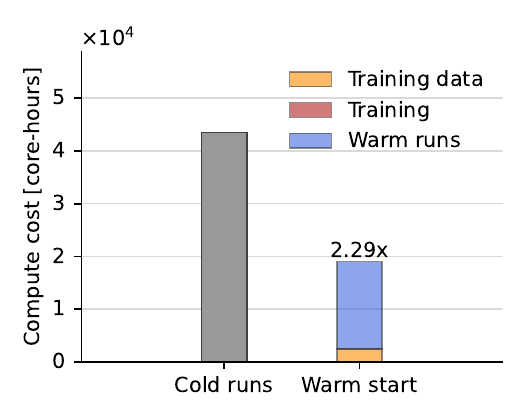}
\caption{CRM effective speed--up}
\end{subfigure}
\begin{subfigure}{0.49\textwidth}
\includegraphics[trim={0 3mm 0 -5mm}, clip, width=0.98\linewidth]{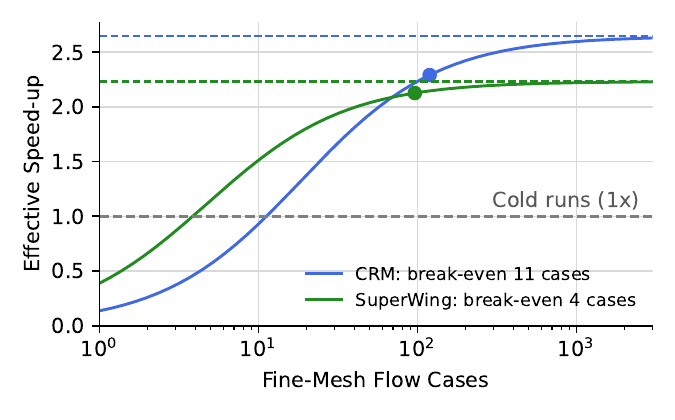}
\caption{Break--even plot for cross--mesh campaigns}
\end{subfigure}
\caption{Campaign cost of cross--mesh deployment using $N=24$ models trained on coarse--mesh data, evaluated at the 1\% convergence criterion. Bars in (a) and (b) compare the cold--start campaign cost with the combined cost of training--data generation, GPU training, and fine--mesh warm--start runs. For cost accounting, one GPU--hour is assigned the same cost as one CPU node--hour (48 core--hours). Panel (c) shows the effective speed--up $S(K)$ defined in Eq.~\eqref{eq:breakeven} as a function of the number of fine--mesh cases. Dots mark the number of cases in respective datasets. Dashed lines indicate the asymptotic limits $\bar{c}_\mathrm{cold}/\bar{c}_\mathrm{warm}$.}\label{fig:effective_speedup}
\end{figure}

The iteration speed--ups in Sections~\ref{sec:superwing}--\ref{sec:f6} exclude the costs of training--data generation and model training. Figure~\ref{fig:effective_speedup} includes these costs for cross--mesh deployment of the $N=24$ models, comparing the combined cost of coarse--mesh training solutions, model training and fine--mesh warm runs with an all--cold fine--mesh campaign. Solver costs are charged up to the iteration at which the 1\% settling criterion is met. GPU training costs correspond to single--precision training on one NVIDIA H100 SXM GPU. For accounting in Fig.~\ref{fig:effective_speedup}(a,b), each GPU-hour is assigned 48 core-hours, equivalent to one CPU node-hour (\ref{app:rungs}). 

For a fixed trained model, let $B$ denote the one--off cost of training--data generation and model training, and let $\bar{c}_\mathrm{cold}$ and $\bar{c}_\mathrm{warm}$ be the mean fine--mesh solver costs per case. A campaign of $K$ cases then costs $K\bar{c}_\mathrm{cold}$ with cold starts and $B+K\bar{c}_\mathrm{warm}$ with learned initialization. The effective speed--up ($S(K)$) and break-even threshold ($K^\ast$) are
\begin{equation}
  S(K)=\frac{K\,\bar{c}_\mathrm{cold}}
  {B+K\,\bar{c}_\mathrm{warm}},
  \qquad
  K^\ast=\frac{B}
  {\bar{c}_\mathrm{cold}-\bar{c}_\mathrm{warm}},
  \label{eq:breakeven}
\end{equation}
where $S(K^\ast)\approx1$, provided warm runs are cheaper on average. Figure~\ref{fig:effective_speedup}(c) varies $K$ while holding these mean costs fixed. As the initial investment is amortized, $S(K)$ approaches the pooled solver--cost ratio $\bar{c}_\mathrm{cold}/\bar{c}_\mathrm{warm}$ shown as dashed lines in respective dataset colors.

Under these assumptions, both cross--mesh campaigns retain more than twofold acceleration after model-building costs are included. SuperWing recovers its smaller initial investment sooner, with break-even estimates of approximately four fine-mesh cases versus eleven for CRM. CRM also approaches its asymptotic speed--up more slowly because training-data generation costs more relative to the saving per fine-mesh case. The initial investment is therefore recovered after only a small fraction of the evaluations required to build a large aerodynamic database\citep{MaueryEtAl2021}.

Favorable amortization depends on both training efficiency and data-generation cost. The H100 provides model--training times of only a few hours, whereas CPU training can cost more than generating the training data. Cross--mesh deployment significantly reduces the initial investment by using inexpensive coarse solutions to accelerate fine--mesh evaluations.  In same--mesh campaigns, additional training cases require converged solutions on the target mesh and increase model--training costs, so higher per--case speed--ups need not yield greater campaign savings.
Under the GPU accounting used here, the most economical same--mesh models use $N=8$--16 (\ref{app:rungs}). Training-set size should therefore balance campaign volume, data--generation and training costs, and the resulting solver savings.

\subsection{Robustness and Multiple Steady States}
\label{sec:robustness}
No warm run diverged. All SuperWing cases satisfy the 1\% convergence criterion on both meshes, while failures are confined to a few CRM and DLR--F6 conditions. 
Among converged runs, only one strongly separated SuperWing fine--mesh case is slower than its cold counterpart. 
\begin{figure}[t]
\centering
\begin{subfigure}{0.49\textwidth}
\includegraphics[trim={2.8mm 2.8mm 2.8mm 5.6mm}, clip, width=0.98\linewidth]{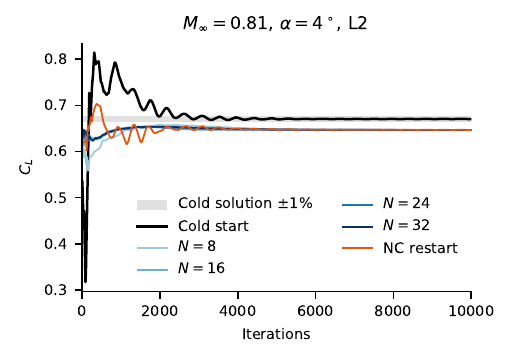}
\caption{$M_\infty=0.81$, $\alpha=4^\circ$, L2}
\end{subfigure}
\begin{subfigure}{0.49\textwidth}
\includegraphics[trim={2.8mm 2.8mm 2.8mm 5.6mm}, clip, width=0.98\linewidth]{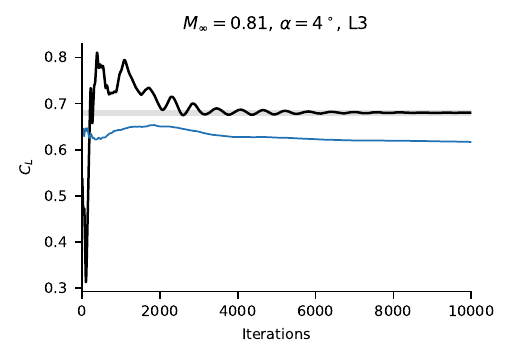}
\caption{$M_\infty=0.81$, $\alpha=4^\circ$, L3}
\end{subfigure}
\begin{subfigure}{0.49\textwidth}
\includegraphics[trim={2.8mm 2.8mm 2.8mm 5.6mm}, clip, width=0.98\linewidth]{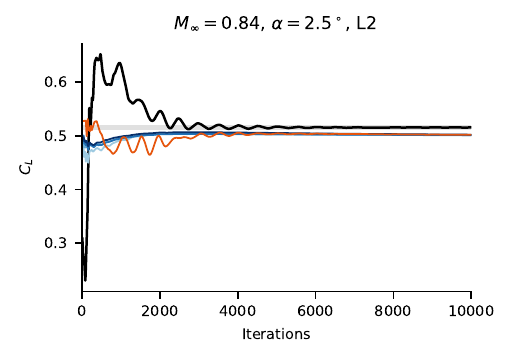}
\caption{$M_\infty=0.84$, $\alpha=2.5^\circ$, L2}
\end{subfigure}
\begin{subfigure}{0.49\textwidth}
\includegraphics[trim={2.8mm 2.8mm 2.8mm 5.6mm}, clip, width=0.98\linewidth]{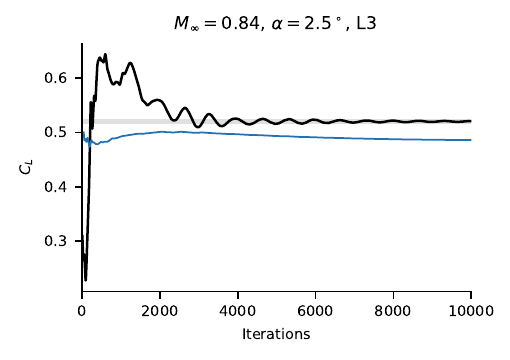}
\caption{$M_\infty=0.84$, $\alpha=2.5^\circ$, L3}
\end{subfigure}
\caption{Lift histories of the two CRM conditions showing initialization-dependent flow states, $M_\infty=0.81$, $\alpha=4^\circ$ (top) and $M_\infty=0.84$, $\alpha=2.5^\circ$ (bottom), at $Re_\infty=5\times10^{6}$, on the training mesh L2 (left) and the finer mesh L3 (right). The cold start is compared with the learned warm starts of the $N=8$--32 models and the NC restart on L2 and of the $N=24$ model on L3; shading: cold solution $\pm1\%$.}
\label{fig:bistable-crm}
\end{figure}

\begin{figure}[t]
  \centering
  \includegraphics[trim={0 3mm 0 7mm}, clip, width=0.56\linewidth]{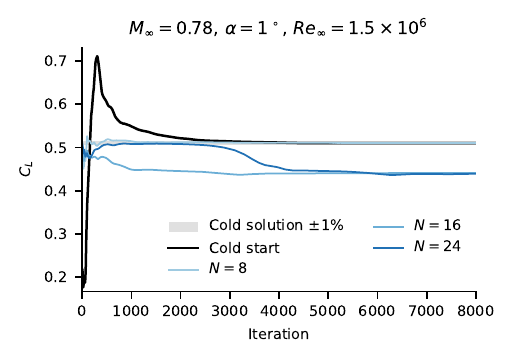}
  \caption{Lift history of the DLR-F6 condition showing initialization-dependent flow states, $M_\infty=0.78$, $\alpha=1^\circ$, $Re_\infty=1.5\times10^{6}$: cold start and learned warm starts of the $N=8$, 16 and 24 models, which end at the 8,000-iteration budget; shading: cold solution $\pm1\%$.}
  \label{fig:bistable-f6}
\end{figure}

The failures highlight sensitivity to initialization near separation onset. On CRM L2, two flow cases with learned warm starts across training--set sizes and NC restarts approach similar lower--lift states with greater wing--body junction separation than the cold solution (Fig.~\ref{fig:bistable-crm}). The discrepancy increases on L3, where the warm histories still drift at the iteration limit. For DLR--F6, a warm run can remain close to the cold lift before departing towards a more separated state (Fig.~\ref{fig:bistable-f6}). Larger training sets and lower aggregate field errors therefore do not guarantee recovery of the cold solution. These observations are consistent with competing RANS solution states near separation onset \citep{KamenetskiyEtAl2014}.

Low residuals or temporary agreement in loads are consequently insufficient to establish convergence to the reference state. Separation--sensitive diagnostics, such as junction pressure or reversed-flow area, should complement the load criterion. 
Alternative initializations and additional training coverage near separation onset may help identify and reduce these discrepancies. The cold solution provides a common assessment reference, but agreement with it alone does not establish physical accuracy; the competing flow states require independent validation.

\subsection{Scalability}
\label{sec:scalability}
Training uses bounded cell and face samples, keeping per-update network computation largely independent of full-mesh resolution. H100 update times vary little across the elevenfold range of training-mesh sizes. With a fixed update budget per case, training time grows approximately linearly with the number of cases, at a projected cost of about 0.1 GPU-hours per case (\ref{app:rungs}). Pre-processing and storage still depend on dataset size. At fixed architecture, inference scales linearly with cell count: physics attention accumulates global slice states across chunks, preserving full-mesh communication without constructing cell-to-cell attention matrices (Section~\ref{sec:training}).

Adopting Transolver--3's physical-state caching and chunkwise decoding could extend the present implementation to significantly larger meshes; the original framework has demonstrated inference over 160\,M cells \citep{ZhouEtAl2026Transolver3}.

Cross--mesh acceleration depends on the correction required between coarse-- and fine--mesh solutions. In the present HiSA configuration, refinement increases cold--start iteration counts. CRM retains or improves its acceleration because the discretization difference is modest, whereas SuperWing's larger solution discrepancy requires more correction and reduces the advantage. These results support coarse--mesh training as a practical way to reduce data-generation cost while accelerating fine--mesh evaluations. Fine--tuning with a few fine--mesh solutions could reduce the remaining discrepancy, although this has not yet been assessed.

\section{Conclusion}
\label{sec:conclusion}
This work developed a Transolver--based warm--start formulation for steady RANS with Spalart--Allmaras and $k$--$\omega$ SST closures. It predicts mean--flow and turbulence fields on native meshes using operating--condition modulation, Fourier and wall features, and an objective tailored to solver correction. Target--gated regularization and relative intercell matching complement field fitting, with case--dependent roughness scales and adaptive gradient--norm balancing.  Prediction requires no preliminary solution or borrowed turbulence fields; training requires neither solver differentiation nor residual evaluation.

Evaluations on SuperWing, CRM and DLR--F6 demonstrate acceleration from small training sets, including off--grid and OOD cases. Generalization across geometry is more demanding than across operating conditions on a fixed configuration, while field accuracy alone does not determine warm--start effectiveness. Using the same 24 CRM solutions, learned initialization achieves a geometric--mean iteration speed--up of $2.34\times$, versus $1.34\times$ for nearest--condition restarts, with more consistent acceleration under the tested pairing rule.

Cross--mesh deployment demonstrates resolution invariance: coarse--mesh models initialize finer meshes without retraining, solution interpolation or target--condition coarse solves. Acceleration depends on the required discretization correction: CRM retains or improves its acceleration benefit, whereas the larger coarse--to--fine discrepancy in SuperWing reduces it. Nevertheless, inexpensive coarse--mesh data enable estimated campaign speed--ups of $2.1\times$ for SuperWing and $2.3\times$ for CRM, including data generation and GPU training. The investment is recovered after approximately four and eleven fine--mesh cases, respectively. 

No warm run diverged, but failures to recover the cold reference in a few cases highlight sensitivity near separation onset. Neither temporary load agreement nor low residuals guarantees reference recovery, and agreement with that reference does not establish physical accuracy. Deployment therefore requires a suitable stopping rule and separation--sensitive monitoring. Additional training coverage near separation onset and limited fine--mesh fine--tuning could improve reliability and reduce discretization discrepancies. Independent initialization, cross--mesh transfer and amortized savings support applications to aerodynamic databases and parallel design evaluations, including cases without suitable neighboring solutions.

\appendix

\section{Model and Training Details}
\label{app:model}

\subsection{Feature and Target Processing}
\label{app:features}

Cell--center coordinates and wall features are obtained from the native CFD mesh. Wall distance is computed using OpenFOAM's \texttt{meshWave} procedure and scaled as defined in Eq.~\eqref{eq:inputs}; wall--normal directions are obtained by normalizing its cell--centered gradient, pointing away from the wall. Coordinates undergo $z$-score standardization using statistics pooled over all training cells. Operating conditions use componentwise statistics across training cases, with unit scale for components of negligible variance. These statistics are held fixed during training and inference. The scaled wall-distance and wall-normal features undergo no further standardization.

For standardized cell--center coordinate $\bar{x}_{ia}$, the Fourier encoding contains
\begin{equation}
 \sin\!\left(\frac{2\pi\sigma_a}{\lambda_k}\bar{x}_{ia}\right),
 \qquad
 \cos\!\left(\frac{2\pi\sigma_a}{\lambda_k}\bar{x}_{ia}\right),
 \qquad
 \lambda_k=\frac{\lambda_0}{2^k},\quad k=0,\ldots,4,
\end{equation}
where $\sigma_a$ is the training coordinate standard deviation and $\lambda_0$ is the median aerodynamic reference length supplied in the training-case metadata. This sets a common physical scale for the five Fourier wavelengths, which remain fixed during training and inference, including cross--mesh deployment.

The turbulence variables $\chi_j\in\{\widetilde{\nu},k,\omega\}$ are transformed according to
\begin{equation}
 \ell_j=\ln\!\left(\frac{\chi_j+f_j}{\chi_{0j}}\right),
 \qquad
 s_{t,j}=\max\!\left(|\ell_{j,\infty}|,1\right).
 \label{eq:app_turbulence}
\end{equation}
Here $\chi_{0j}$ is a unit reference quantity that makes the logarithm's argument dimensionless: $1\,\mathrm{m^2\,s^{-1}}$ for $\widetilde{\nu}$, $1\,\mathrm{m^2\,s^{-2}}$ for $k$, and $1\,\mathrm{s^{-1}}$ for $\omega$. The positive offsets are
$f_{\widetilde{\nu}}=10^{-9}\,\mathrm{m^2\,s^{-1}}$,
$f_k=10^{-6}\,\mathrm{m^2\,s^{-2}}$, and
$f_\omega=1\,\mathrm{s^{-1}}$.
The freestream value $\ell_{j,\infty}$ is obtained by applying the same logarithmic transformation to the prescribed initial freestream turbulence value. The resulting channels are normalized through Eq.~\eqref{eq:targets}; no additional variance normalization is applied to the outputs.

A binary validity mask $v_i$ assigns zero loss weight to cells with nonfinite target values, nonpositive pressure, negative $\widetilde{\nu}$ or $k$, nonpositive $\omega$, or
\begin{equation}
 T_i>1.02T_{0,\infty},
 \qquad
 T_{0,\infty}=T_\infty
 \left[1+\frac{\gamma-1}{2}M_\infty^2\right],
 \qquad \gamma=1.4.
\end{equation}
The temperature criterion screens localized numerical overshoots. The implemented criterion uses static temperature. 

\subsection{Network Configuration and Optimization}
\label{app:architecture}
The feature embedding maps the seven mesh features $(\mathbf{x},d_w,\mathbf{n}_w)$, augmented by 30 coordinate features from Fourier encoding, through a $37\to640\to320$ MLP. The three preprocessed operating conditions are mapped through a $3\to320\to320\to320$ encoder. The network contains eight
Transolver blocks, each with feature width $D=320$, ten attention heads, 64 slice tokens per head, and a
$320\to640\to320$ MLP. Slicing uses deterministic soft assignments during both training and inference, and dropout is disabled.

Each Transolver block applies physics attention followed by a cell--wise MLP, with separate scale-adjustment, shift and gate vectors derived from the operating-condition encoding for each update. Incoming features are layer-normalized across channels at each cell, scaled and shifted, then passed through the corresponding operation. Its output is gated componentwise and added to the features entering that update through a residual connection. The modulation vectors are specific to each block and update but shared across all cells, since they depend only on the global operating conditions.

The perturbation decoder applies layer normalization with condition-dependent scaling and shifting, followed by a $320\to C$ linear projection. Here $C=6$ for SA and $C=7$ for SST, corresponding to the normalized target channels defined in Eq.~\eqref{eq:targets}.

Training uses AdamW with a maximum learning rate of $5\times10^{-4}$, weight decay $10^{-4}$, and gradient-norm clipping at one. A one-cycle schedule assigns ten percent of the updates to warm-up, with initial and nominal final learning rates of $2\times10^{-5}$ and $2\times10^{-8}$, respectively. Each epoch samples the training cases in shuffled order and performs five updates per case using fresh samples. SuperWing and CRM use 140 epochs, corresponding to 700 updates per case; DLR-F6 uses 175 epochs, corresponding to 875 updates per case. All warm starts use final-epoch model parameters. Full-mesh inference uses the global slice accumulation described in Section~\ref{sec:training}.

\subsection{Loss Weights and Sampling}
\label{app:objective}

The loss averages in Eq.~\eqref{eq:loss_terms} are arithmetic
means over sampled entries and output channels, without
cell-volume or face-area weighting. Masked entries contribute
zero without changing the averaging denominator.

Let $z_{ic}$ denote the dimensionless target perturbation
at cell $i$ in channel $c$. The first five of the $C$ output
channels represent $(U_x,U_y,U_z,p,T)$, followed by turbulence.
For internal face $e$, connecting owner cell $o(e)$ and
neighbour cell $n(e)$, define
$\Delta_e z_c=z_{o(e),c}-z_{n(e),c}$.
The binary mask $v_i$ equals one for targets passing the
validity checks in \ref{app:features} and zero otherwise;
$v_e$ equals one only when both adjacent cells are valid.
The scaled wall distance $d_i\in[0,1)$ increases away
from the wall, and $b_i$ identifies cells owning an
aerodynamic wall face. Base channel weights are $w_c=2$
for velocity components and pressure and $w_c=1$ for
temperature and turbulence.

Define the wall-distance ramp $R_a$, mean-flow departure
$D_i$, target face roughness $r_e$ and smoothness gate $g_e$ as
\begin{equation}
\begin{aligned}
R_a(d)
&=\left[\operatorname{clip}
  \left(\frac{d-a}{1-a},0,1\right)\right]^2,
\qquad
D_i=\frac{1}{5}\sum_{c=1}^{5}w_c|z_{ic}|,\\
r_e
&=\frac{1}{C}\sum_{c=1}^{C}w_c|\Delta_e z_c|,
\qquad
g_e=R_{0.50}
\left(\frac{d_{o(e)}+d_{n(e)}}{2}\right)
\exp(-r_e/\tau_r).
\end{aligned}
\end{equation}
Here $a$ sets the ramp threshold, and
$\operatorname{clip}(x,0,1)$ restricts $x$ to $[0,1]$.
For $g_e$, the wall-distance factor $R_{0.50}$ suppresses smoothing
near walls, while the exponential factor reduces it across
sharp target variations, helping preserve shocks and shear layers.

The data, freestream, smoothness and jump-matching weights are
\begin{equation}
\begin{aligned}
w^{\mathrm d}_{ic}
&=v_i[1+5(1-d_i)](1+2b_i)w_c,\\
w^{\mathrm f}_i
&=v_iR_{0.90}(d_i)\exp[-(D_i/0.02)^2],\\
w^{\mathrm s}_e
&=v_eg_e,
\qquad
w^{\mathrm j}_{ec}=h_ev_e(1-g_e)\widetilde w_c.
\end{aligned}
\label{eq:app_loss_weights}
\end{equation}
Here $h_e$ accounts for nonuniform face sampling, while
$\widetilde w_c$ emphasizes pressure and temperature
and assigns zero weight to turbulence channels.
The complementary gate $1-g_e$ emphasizes jump matching
where smoothing is reduced. Freestream regularization
acts on all channels, although its gate uses only
mean-flow targets.

For each case, the roughness scale $\tau_r$ and the
denominator floor $\varepsilon_{\mathrm j}$ in the
jump-matching loss of Eq.~\eqref{eq:loss_terms} are set
from the 70th and 50th percentiles of $r_e$, respectively.
Both are used only during training.

\section{Dataset Construction and Partitioning}
\label{app:datasets}
\subsection{Mesh Sources and Reference-Solution Screening} SuperWing meshes are generated using a Python implementation adapted from the published meshing procedure~\citep{YangEtAl2025SuperWing}, combining structured surface meshing with hyperbolic volume extrusion using \texttt{pyHyp}. The original ADflow study assessed this procedure through a three-resolution grid-convergence study on the CRM wing~\citep{YangEtAl2025SuperWing}. For the present study, the coarse meshes are obtained from fine meshes by halving the resolution in each computational direction. Both levels represent the same geometries and operating conditions.

The CRM meshes are the coarse L2 and medium L3 hexahedral meshes from the Drag Prediction Workshop(DPW)-5 common wing--body grid family,\footnote{\url{https://dpw.larc.nasa.gov/DPW5/}} developed by \citet{Vassberg2011UnifiedGrid}. This family formed the basis of the workshop's systematic grid-refinement and inter-code comparisons~\citep{LevyEtAl2014}. The downloaded meshes are converted to OpenFOAM format for the present simulations.

The DLR-F6 mesh is the coarse ICEM-CFD point-matched multiblock mesh for the wing--body--nacelle--pylon configuration, obtained from the DPW-2 archive.\footnote{\url{https://dpw.larc.nasa.gov/DPW2/b2b-icem/}} Its hexahedral blocks include the flow-through nacelle passage and are converted to OpenFOAM format. The mesh belongs to the benchmark collection used for grid-refinement studies and comparisons with wind-tunnel measurements in DPW-2~\citep{LaflinEtAl2005}. 

Before generating the extended datasets, the mesh conversions and CFD case setups were assessed by comparing baseline simulation results with the available reference solutions. The cell-level masks in~\ref{app:features} limit the training influence of localized field anomalies; they neither replace this convergence assessment nor modify the reference force histories.

\subsection{CFD Setup}
\label{app:cfd_setup}

All datasets use steady, compressible, fully turbulent RANS simulations on half-model domains with symmetry boundaries, no-slip adiabatic walls, and characteristic farfield conditions. Air is represented by a perfect-gas equation of state with Sutherland viscosity. The freestream temperatures are 300 K for SuperWing, 310.928 K for CRM, and 305.06 K for DLR-F6. Reynolds numbers use reference lengths of 1 m, 7.00532 m, and 0.1412 m, respectively. At fixed temperature, density is adjusted to obtain the prescribed Reynolds number.

The HiSA discretization uses AUSMPlusUp fluxes, \texttt{wVanLeer} reconstruction, and limited least-squares gradients. Turbulence convection uses a bounded limited-linear scheme. The coupled flow equations are advanced in pseudo-time using GMRES with LU-SGS preconditioning. DLR-F6 uses a cubic gradient limiter, whereas CRM and SuperWing use the standard cell limiter. Maximum pseudo-Courant numbers are 5, 10, and 25 for CRM, DLR-F6, and SuperWing, respectively. Numerical settings are retained within each paired cold- and warm-start comparison.

\subsection{Training and Evaluation Partitions}
\label{app:partitions}
Training-case selection combines randomization with balanced representation of operating conditions. The aim is a reproducible rule that can be applied to an eligible case pool without retrospective adjustment based on prediction errors or warm-start performance. After canonical sorting, case groups are visited in a seeded random order. Within each group, selection favors the parameter level least represented among the cases selected so far. The resulting ordering is frozen, and its first $N_{\mathrm{train}}$ entries define each training subset.

For SuperWing, each group is a geometry, and two of its three angles of attack are selected using the same balancing principle. Training-set sizes of 8, 16, 24, and 32 cases therefore correspond to 4, 8, 12, and 16 trained geometries. The third angle remains held out: withholding $7^\circ$ tests interpolation in incidence, whereas withholding $6^\circ$ or $8^\circ$ tests extrapolation relative to that geometry's training angles. Entirely untrained geometries provide a separate assessment of geometric generalization.

For CRM and DLR-F6, each group comprises a Mach--AoA pair, with members differing in Reynolds number. Groups are picked randomly and one case is selected per group on each pass, balancing Reynolds-number representation as the training subset grows. Off--grid and out--of--range cases remain exclusively for evaluation.

\section{Solver Restart and Assessment Details}
\label{app:protocol}

\subsection{Restart Construction and Solver Controls}
\label{app:restart}
The predicted perturbations are first converted to dimensional mean-flow variables and logarithmic turbulence components $\widehat{\boldsymbol\ell}_{t,i}$ by reversing the Eq.~\eqref{eq:targets}. For turbulence component $j$, the inverse of Eq.~\eqref{eq:app_turbulence} gives
\begin{equation}
 \widehat{\chi}_{i,j}
 =
 \max\!\left(
 \chi_{0j}\exp(\widehat{\ell}_{t,i,j})-f_j,\,
 \chi_{\min,j}
 \right),
\end{equation}
where $\widehat{\ell}_{t,i,j}$ is component $j$ of
$\widehat{\boldsymbol\ell}_{t,i}$. The internal-cell minima are
$\widetilde{\nu}_{\min}=10^{-9}\,\mathrm{m^2\,s^{-1}}$,
$k_{\min}=0$, and $\omega_{\min}=1\,\mathrm{s^{-1}}$.
Boundary conditions remain unchanged. Export applies no pressure or
temperature clipping, freestream blending, or conservative projection.

The OpenFOAM interface writes $\mathbf{U},p,T,\widetilde{\nu}$ for SA closure model, or $\mathbf{U},p,T,k,\omega$ for SST closure model. Predicted internal fields are inserted into field--file templates while retaining their boundary-condition entries. The prepared warm case starts at time zero with the original mesh and numerical dictionaries. The auxiliary fields \texttt{nut} and \texttt{alphat} are carried from the cold initialization; they are not neural outputs and are updated by the active turbulence and transport models.

For the paired HiSA comparisons, the cold run's residual-reference metadata are copied through \texttt{pseudotimeState}. This retains the reference used by the pseudo-time controller instead of redefining it from the predicted initial field. This restart convention is separate from the neural prediction and is part of the fair comparison protocol. Spatial schemes, turbulence closure, and pseudo-time control settings otherwise remain those of the corresponding cold case.

\subsection{Convergence Measures}
\label{app:convergence}
Initial--field accuracy compares the neural prediction with the converged cold-run field before solver correction. For case $a$, $\widehat{\mathbf Z}_a$ and $\mathbf Z_a$ contain the predicted and reference perturbations of Eq.~\eqref{eq:targets} on the same mesh. The relative error and its mean over $N_\mathrm{eval}$ cases are
\begin{equation}
  e_a=
  \frac{\lVert\widehat{\mathbf Z}_a-\mathbf Z_a\rVert_F}
       {\lVert\mathbf Z_a\rVert_F+10^{-8}},
  \qquad
  E=\frac{1}{N_\mathrm{eval}}\sum_{a=1}^{N_\mathrm{eval}}e_a .
  \label{eq:field-error}
\end{equation}
The norm includes all cells and normalized channels without cell-volume weighting; turbulence channels remain logarithmic. Solver convergence is assessed separately using the total aerodynamic coefficients.

\paragraph{Reference values and tolerance band}
For each $C\in\{C_L,C_D,C_M\}$, $C_\mathrm{ref}$ and $\sigma_C$ are the mean and standard deviation of the final 200 iterations of the long cold run. Both initializations use the same reference and relative tolerance, with band half-width
\begin{equation}
  b_C=\max\bigl(\epsilon\,|C_\mathrm{ref}|,\;b_{C,\min}\bigr),
  \label{eq:band}
\end{equation}
where $\epsilon=0.01$ for the main assessment. The absolute floor is $b_{C,\min}=0.001$ for lift and zero otherwise. Bands are resolvable when $b_C>2\sigma_C$. Sensitivity to $\epsilon=0.005$ and $0.02$ is reported in \ref{app:tolerance}.


\paragraph{Failures and final-state checks}
Settling iterations follow the criterion defined in Section~\ref{sec:protocol}. Warm runs that do not meet the criterion within their iteration budget count as failures. Terminal differences are measured by
\begin{equation}
\Delta C=
\frac{\overline{C}_\mathrm{warm}-\overline{C}_\mathrm{cold}}
   {|\overline{C}_\mathrm{cold}|},
\label{eq:final-gap}
\end{equation}
using final-200-iteration means. Surface checks additionally compare the wetted-area fraction with negative streamwise skin friction. Failed runs are excluded from speed--up averages and discussed in \ref{sec:robustness}.



\section{Supplementary Results}
\label{app:supplementary}

\subsection{Sensitivity to the Convergence Tolerance}
\label{app:tolerance}

Table~\ref{tab:tolerance} repeats the per-case summaries of Figs.~\ref{fig:sw_speedup}--\ref{fig:f6_speedup} at $\epsilon=0.5\%$ and 2\%. Cold and warm runs are scored at the same tolerance, so both iteration counts change, whereas the reference values remain the means of the long cold runs. Over all non-training cases, the speed-ups at 0.5\% are 7--16\% lower than at 1\% for the CRM and SuperWing and 8--32\% lower for DLR-F6; at 2\%, they are 7--24\% and 8--14\% higher, respectively. The SuperWing Train values on L1 are the most sensitive, 24--31\% lower at 0.5\%, because these warm runs settle after only 250--380 iterations (median) at 1\%. For the non-training categories, the trends with $N$ are unchanged, except for the two DLR-F6 OOD conditions, whose cold runs settle only after 6,700--7,700 iterations at 0.5\%. The cross-mesh comparisons are also unchanged: on L3, the CRM speed-up exceeds the L2 value in every category at 1\% and 2\% and in three of the four at 0.5\%, and the SuperWing speed-up on L2 remains below the L1 value in every category at every tolerance. The failures are nearly independent of the tolerance.
\begin{table}[t]
\centering
\caption{Geometric--mean speed--up $N_\mathrm{cold}/N_\mathrm{warm}$ at three tolerances, cold and warm runs being scored at the same $\epsilon$. Tr: Train; Te: Test; Off: Off-grid.}
\label{tab:tolerance}
\footnotesize
\setlength{\tabcolsep}{3.2pt}
\begin{tabular}{@{}llcccccccccccc@{}}
\toprule
 & & \multicolumn{4}{c}{$\epsilon=0.5\%$} & \multicolumn{4}{c}{$\epsilon=1\%$} & \multicolumn{4}{c}{$\epsilon=2\%$}\\
\cmidrule(lr){3-6}\cmidrule(lr){7-10}\cmidrule(lr){11-14}
Dataset (mesh) & $N$ & Tr & Te & Off & OOD & Tr & Te & Off & OOD & Tr & Te & Off & OOD\\
\midrule
SuperWing (L1) & 8  & 3.05 & 1.97 & -- & 1.90 & 4.00 & 2.15 & -- & 2.12 & 5.57 & 2.51 & -- & 2.36\\
               & 16 & 3.44 & 2.17 & -- & 2.17 & 4.97 & 2.45 & -- & 2.50 & 5.64 & 2.77 & -- & 2.71\\
               & 24 & 3.47 & 2.40 & -- & 2.50 & 5.02 & 2.82 & -- & 2.81 & 5.87 & 3.31 & -- & 3.35\\
               & 32 & 3.43 & 2.61 & -- & 2.68 & 4.77 & 3.09 & -- & 3.18 & 5.97 & 3.97 & -- & 3.59\\
SuperWing (L2)$^\dagger$ & 24 & 2.15 & 1.91 & -- & 2.04 & 2.48 & 2.18 & -- & 2.68 & 3.22 & 2.65 & -- & 3.06\\
\addlinespace
CRM (L2)       & 8  & 2.17 & 1.86 & 1.89 & 1.84 & 2.56 & 2.00 & 1.95 & 1.93 & 2.96 & 2.14 & 2.11 & 2.10\\
               & 16 & 2.34 & 2.07 & 2.03 & 1.97 & 2.60 & 2.26 & 2.14 & 2.29 & 3.05 & 2.55 & 2.30 & 2.69\\
               & 24 & 2.38 & 2.16 & 2.09 & 2.17 & 2.74 & 2.37 & 2.17 & 2.51 & 2.87 & 2.63 & 2.34 & 2.96\\
               & 32 & 2.54 & 2.27 & 2.23 & 2.50 & 2.93 & 2.49 & 2.29 & 2.91 & 3.19 & 2.65 & 2.49 & 3.06\\
CRM (L3)$^\dagger$ & 24 & 2.47 & 2.28 & 2.32 & 2.08 & 2.82 & 2.55 & 2.61 & 2.63 & 3.16 & 2.80 & 2.74 & 3.31\\
\addlinespace
DLR-F6         & 8  & 2.49 & 2.31 & 2.42 & 1.46 & 2.52 & 2.43 & 2.77 & 2.23 & 2.65 & 2.59 & 2.87 & 3.26\\
               & 16 & 2.98 & 2.62 & 2.68 & 1.45 & 3.22 & 3.55 & 3.89 & 4.89 & 4.26 & 3.83 & 4.50 & 5.37\\
               & 24 & 3.06 & 3.29 & 3.05 & 2.06 & 3.68 & 3.73 & 4.11 & 5.09 & 4.21 & 4.40 & 4.59 & 4.91\\
\bottomrule
\end{tabular}

\smallskip
\begin{minipage}{\linewidth}
\footnotesize
\qquad \quad \, $^\dagger$Cross-mesh inference with the $N=24$ model trained on the coarser mesh.
\end{minipage}
\end{table}

\subsection{Training Time and Memory Requirements}
\label{app:rungs}
Table~\ref{tab:training-cost} summarizes the training requirements
of representative models. Each epoch performs five updates per
training case. The estimated training times are a few hours on
one GPU, compared with several days on the CPU configuration,
supporting GPU training to reduce the initial investment in
learned warm starts.
\begin{table}[t]
\centering
\small
\caption{Training budgets and estimated total training times
in single precision, using 24 threads of an AMD EPYC 7352 CPU
or one NVIDIA H100 SXM GPU (80\,GB). Times are wall-clock hours (h).}
\label{tab:training-cost}
\setlength{\tabcolsep}{5pt}
\begin{tabular}{@{}lcccccc@{}}
\toprule
Dataset & $N$ & Epochs &
\shortstack{GPU time\,(h)} &
\shortstack{CPU time\,(h)} &
\shortstack{GPU peak\,(GiB)} \\
\midrule
SuperWing L1 & 24 & 140 & 1.93 & 132.7 & 43.9 \\
CRM L2       & 24 & 140 & 2.33 & 195.5 & 54.1 \\
DLR--F6      & 16 & 175 & 2.00 & 163.3 & 56.0 \\
\bottomrule
\end{tabular}
\end{table}

\subsection{Nearest-Condition Restarts}
\label{app:nc-restart}

The NC baseline initializes each of the 97 non-training CRM conditions from one of the 24 training solutions. The pairing rule, fixed before running the comparisons, selects the source minimizing
\begin{equation}
d=\left[
\left(\frac{\Delta M_\infty}{0.27}\right)^2+
\left(\frac{\Delta\alpha}{5.5^\circ}\right)^2+
\left(\frac{\Delta\log_{10}Re_\infty}{\log_{10}7}\right)^2
\right]^{1/2},
\label{eq:nc-distance}
\end{equation}
where the denominators are the nominal parameter spans. Ties favour lower source incidence, followed by smaller Reynolds-number and Mach-number differences.

Source internal fields are transferred using
\begin{equation}
p_\mathrm{t}=p_\mathrm{s}
\frac{p_{\infty,\mathrm{t}}}{p_{\infty,\mathrm{s}}},
\qquad
\tilde{\nu}_\mathrm{t}=\tilde{\nu}_\mathrm{s}
\frac{\tilde{\nu}_{\infty,\mathrm{t}}}{\tilde{\nu}_{\infty,\mathrm{s}}},
\qquad
\mathbf{U}_\mathrm{t}=\mathbf{U}_\mathrm{s},
\qquad
T_\mathrm{t}=T_\mathrm{s},
\end{equation}
where s and t denote source and target. With common freestream temperature, this preserves pressure, velocity, temperature and turbulence fields normalized by their respective freestream pressure, sound speed, temperature and turbulence-variable scales. Target boundary conditions are retained, and field export, iteration budgets, numerical controls, core counts and convergence assessment match those of the learned warm starts.

The reported comparison uses the 95 cases satisfying the common cold-reference criterion; two cases fail this criterion with both initializations (Section~\ref{sec:robustness}). The aggregate advantage of learned initialization also persists at the 0.5\% and 2\% tolerances. No distance weights or transfer rules were tuned to restart performance, and alternative source-selection strategies were not evaluated.

\bibliographystyle{elsarticle-harv} 
\bibliography{references}

\end{document}